\documentclass[10pt,twocolumn,letterpaper]{article}

\usepackage[spanish,english]{babel}
\usepackage[utf8x]{inputenc}
\usepackage[T1]{fontenc}
\usepackage{tikz}
\usepackage{float}
\usepackage{subcaption}
\usepackage{wrapfig}
\usepackage{multirow}

\usepackage{booktabs}
\usepackage[normalem]{ulem}
\useunder{\uline}{\ul}{}

\usepackage[a4paper,top=1cm,bottom=2cm,left=1cm,right=1cm,marginparwidth=1.75cm]{geometry}

\usepackage{amsmath}
\usepackage{graphicx}
\usepackage[colorlinks=true, allcolors=blue]{hyperref}

\title{
\usefont{OT1}{bch}{b}{n}
\normalfont \normalsize \textsc{Purdue CS53600 - Data Communication and Computer Networks} \\ [10pt]
\huge RouteNet-Fermi: Network Modeling With GNN \\ Analysis And Re-implementation
}

\usepackage{authblk}

\author{Shourya Verma}
\author{Simran Kadadi}
\author{Swathi Jayaprakash}
\author{Arpan Kumar Mahapatra}
\author{Ishaan Jain}

\affil{Department of Computer Science, Purdue University}

\begin{document}
\maketitle

\selectlanguage{english}

\normalsize
\section{Abstract}
Network performance modeling presents important challenges in modern computer networks due to increasing complexity, scale, and diverse traffic patterns. While traditional approaches like queuing theory and packet-level simulation have served as foundational tools, they face limitations in modeling complex traffic behaviors and scaling to large networks. This project presents an extended implementation of RouteNet-Fermi, a Graph Neural Network (GNN) architecture designed for network performance prediction, with additional recurrent neural network variants. We improve the the original architecture by implementing Long Short-Term Memory (LSTM) cells and Recurrent Neural Network (RNN) cells alongside the existing Gated Recurrent Unit (GRU) cells implementation. This work contributes to the understanding of recurrent neural architectures in GNN-based network modeling and provides a flexible framework for future experimentation with different cell types. \\ \\
{\textbf{Keywords:} \\
Network Modeling, Graph Neural Networks} \\ \\
\href{https://github.com/shouryaverma/routenet_fermi}{Code on Github}\\
\href{https://medium.com/@shourya.verma11/network-modeling-with-gnns-using-routenet-fermi-re-implementation-and-analysis-ad96df395d48}{Blog on Medium}\\
\href{https://youtu.be/jhcQ55bDzSc}{Video on Youtube}

\section{Introduction}

The increasing complexity and scale of modern computer networks have created a need for modeling tools that are accurate and efficient. These tools are useful for network planning, optimization, and management tasks, but  traditional approaches face many limitations in addressing current challenges. Network operators and researchers are trying to develop models that can accurately predict performance metrics across diverse network configurations, traffic patterns, and scheduling policies while maintaining computational efficiency and scalability. Traditional network modeling approaches, primarily queuing theory and packet-level simulation, have served as the backbone of network analysis for decades. Queuing theory provides analytical models that are computationally efficient but relies on restrictive assumptions about traffic distributions, assuming Markovian arrival processes that poorly represent real-world network traffic. Whereas packet-level simulators offer high accuracy but become computationally expensive when modeling large-scale networks or processing high-volume traffic. Recent advances in deep learning, particularly in the domain of Graph Neural Networks (GNNs), have shown new methods for network modeling. RouteNet-Fermi, introduced by Ferriol-Galmés et al. \cite{3}, is a contribution in this direction, demonstrating how GNNs can effectively model complex network behaviors while maintaining computational efficiency. Their architecture employs a three-stage message-passing algorithm that captures the important relationships between network flows, queues, and links, using Gated Recurrent Units (GRUs) to process sequential network state information. Our work extends the original RouteNet-Fermi architecture by implementing and evaluating additional Recurrent Neural Network (RNN) cell variants. Specifically, we introduce Long Short-Term Memory (LSTM) cell and simple RNN cell implementations alongside the existing GRU cells. This extension is motivated to explore the current state of the art machine learning methods in sequence modeling that suggests different RNN architectures may offer varying advantages in terms of memory retention, gradient flow, and computational efficiency. The flexibility of our implementation allows for direct comparison of different recurrent architectures under identical network conditions and same hyper-parameters, providing insights into their relative strengths and weaknesses.

\section{Background}

\subsection{Traditional Approaches}

Network modeling has been essential in the scaling and optimization of large computer networks. By understanding the performance metrics such as delay, jitter, and packet loss, these network models allows researchers to simulate the infrastructure improving system reliability and helps increase performance \cite{2}. Network modeling has many applications such as, protocol design, capacity planning, traffic engineering, and Quality of Service (QoS) optimization. Existing network modeling methods, such as Queuing Theory (QT) and simulation-based tools have been used to understand the way networks behave and perform \cite{1}. QT, with its abstract foundation, gives an organized framework for calculating performance metrics, based on assumptions such as Poisson arrivals and Markovian processes. These assumptions removes the complexity from analysis, making it more manageable, but they limit the applicability of QT in real-world scenarios where traffic patterns often show heavy tails, autocorrelation, and bursts \cite{4,5}. similarly, QT has difficulty capturing the complex dependencies and unpredictable variability characteristic of modern network traffic. Simulation-based tools like OMNeT++ \cite{6} and NS-3 \cite{7} control these obstacle by offering high speed, packet-level simulations of network components. These tools are useful at capturing delicate details and assessing network performance under many conditions. so, their measurable needs grows proportional with the size of the traffic intensity and network, making them not feasible for large-scale networks or when requiring quick results \cite{8}. To overcome this limitation, we use different topologies with OMNET++ simulator which will calculate the delay in source-destination flows. The network topologies are generated using an algorithm called power-Law Out-degree algorithm \cite{9}. Similarly, these simulators face difficulties to generalize results across different network topologies, traffic patterns, and configurations, specifying the need for more adaptable and adjustable techniques.

The challenges in the traditional methods are seen in modern networking framework that has non-Markovian traffic patterns, highly dynamic topologies, and diverse QoS policies \cite{19,20}. Networks with various requirements, such as those in 5G and more, need advanced modeling frameworks that is capable of adapting to the growing needs while not compromising on the computational efficiency \cite{21}. Recent innovations in machine learning have brought many innovative solutions. Graph Neural Networks (GNNs) have come out to be one of the powerful tool for modeling the network \cite{12,14}. Usually traditional neural networks work on Euclidean data, but, GNNs are designed to work with graph-structured data, which describes the network topologies with nodes (e.g., devices) and edges (e.g., links) \cite{15,16}. By using message-passing technology, GNNs capture complicated relationships among queues, process and links, which allows them to model more intricate networks. This ability to learn from graph-based representations makes GNNs acceptable for network performance prediction, ensuring adaptability at scale and offering flexibility \cite{22}.

\subsection{RouteNet-Fermi Framework}

\begin{figure}[t]
    \centering
    \includegraphics[width=0.9\linewidth]{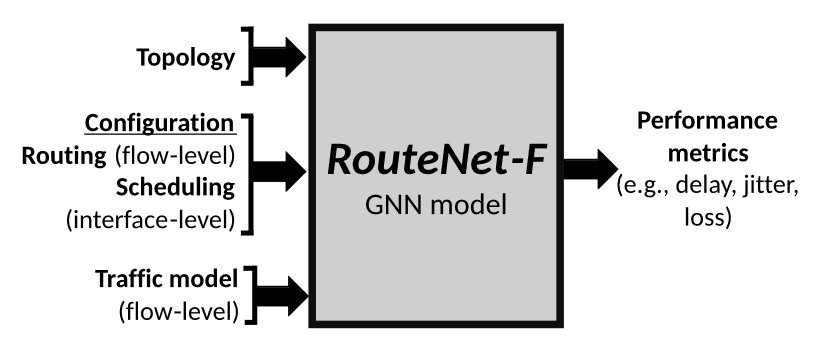}
    \caption{Black-box representation of RouteNet-Fermi}
    \label{fig:routenet}
\end{figure}

\begin{figure}[t]
    \centering
    \includegraphics[width=0.9\linewidth]{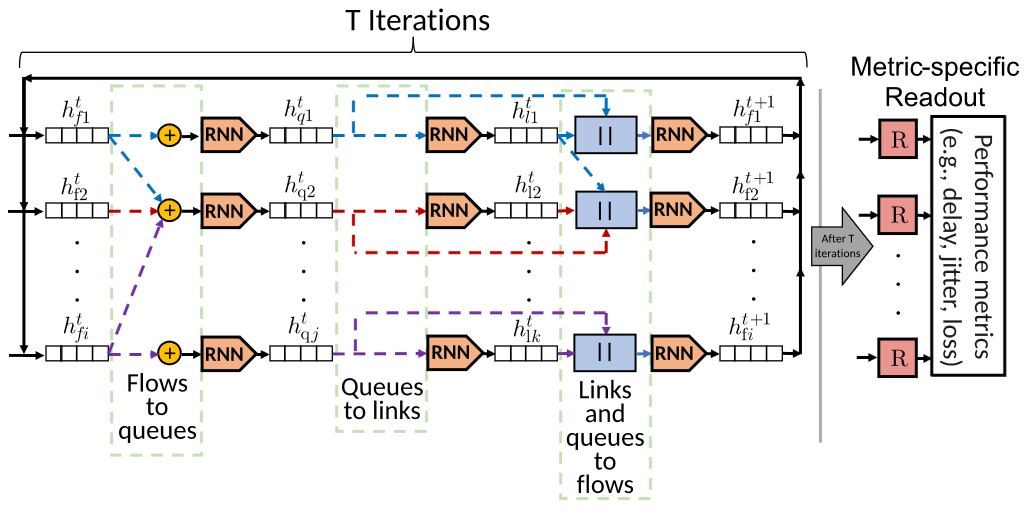}
    \caption{Schematic representation of RouteNet-Fermi}
    \label{fig:schematic}
\end{figure}

RouteNet-Fermi stands out as a state-of-the-art GNN-based architecture especially designed for modeling the network. Figure \ref{fig:routenet} shows a black-box representation of RouteNet-Fermi. The input of this model is a network sample, defined by: a network topology, a routing scheme (flow level), a queuing configuration (interface level), and a set of traffic flows characterized by some parameters. As output, the model produces estimates of relevant performance metrics at a flow-level granularity (e.g., delay, jitter, packet loss). RouteNet-Fermi uses a three step message passing algorithm which finds the dependencies between queues, flows and links, ensuring precise prediction of key performance metrics like delay, jitter, and packet loss. Figure \ref{fig:schematic} shows a schematic representation of the internal three-stage message-passing architecture of this model. This paper talks about the challenges in scalability, generalization, efficiency and illustrates how to replace traditional simulators without compromising on accuracy \cite{22}. 

The model's architecture is built on two fundamental design principles: finding effective representations of network components and exploiting scale-independent features to handle larger networks not seen during training. The network is formally defined as sets of flows, queues, and links, where flows are represented as sequences of queue-link tuples along their source-destination paths. The model maintains state information about these components through hidden state vectors that capture the complex features between them. These relationships are mathematically formalized through graph functions that operate on the hidden states of flows, queues, and links respectively. Initial hidden states are created using specialized functions that encode input features into fixed-size vectors. The flow features include traffic volume and model-specific parameters, link features comprise load information and scheduling policies, and queue features contain buffer sizes and priority levels. For scaling to larger networks, instead of using raw capacity values, it works with relative ratios and predicts delays indirectly through queue occupancy. This approach helps maintain accuracy when dealing with networks larger than those seen during training.

Our research builds upon the foundation of RouteNet-Fermi by enhancing its architecture with additional recurrent components. While the original framework employed Gated Recurrent Units (GRUs) \cite{10,11} for continuous processing, we incorporate Long Short-Term Memory (LSTM) cells and simple RNN cells \cite{10} into the model. LSTMs, with advanced gating procedure and explicit memory cells, are appropriate for finding the long term dependencies in the development of network state. But RNN cells provide computational efficiency, which are powerful where short-term dependencies dominate. The objective is to find the effect of various recurrent architectures on GNN-based network design. This require examining the options between memory retention, computational complexity, gradient flow, and predictive accuracy across various network scenarios. End-to-end delay prediction necessitates understanding congestion dynamics and traffic models, requiring the model to balance queuing and transmission delays. Jitter prediction, which is highly sensitive to variations in packet delays, pushes the recurrent cells' capacity to capture temporal dependencies and burstiness in traffic. Finally, packet loss prediction demands nuanced insights into buffer management and queue overflow patterns, which depend heavily on the scheduling policies. By implementing a unified framework that uses these three recurrent variants, our contribution allows for a side-by-side evaluation of their strengths and weaknesses under identical experimental conditions. This flexibility advances the understanding of GNN-based network models and also offers insights for future research. Our framework’s adaptability ensure its relevance across diverse network scenarios, making it a step forward in data-driven network modeling.

\section{Dataset}
We evaluated our extended RouteNet-Fermi model using a set of datasets generated using the OMNeT++ network simulator, v5.5.1 \cite{23}. An image of the simulator is publicly available \cite{24}. The datasets are categorized into three main groups: scheduling policies, scalability, and traffic profiles. Each category tests specific aspects of the model's performance, enabling a thorough evaluation across different network conditions.

\subsection{Scheduling Policies Datasets}
To assess the impact of different queue scheduling mechanisms on network performance, we used datasets implementing FIFO, Strict Priority (SP), Weighted Fair Queuing (WFQ), and Deficit Round Robin (DRR) policies—where all except FIFO utilize three queues with varying priorities at each node. For WFQ and DRR, multiple weight configurations were defined, and both scheduling policies and buffer sizes were randomly selected per node but remained consistent across all its ports. In the Fat Tree datasets, we employed Fat Tree topologies with 16, 64, and 128 nodes, following the parameters and specifications described in \cite{25}, each with a single routing configuration using FIFO scheduling and a queue size of 32,000 bits. The traffic profile included Poisson time distributions and a binomial packet size distribution between 300 and 1,700 bits, with traffic randomly distributed among all paths based on a selected traffic intensity (TI) to simulate realistic network conditions.

Additionally, in the Scheduling dataset, we examined the model's ability to handle different scheduling policies within the same network by using NSFNET \cite{26} (14 nodes) and GEANT \cite{27} (24 nodes) topologies for training and the GBN \cite{28} (17 nodes) topology for testing, each with 100 variations of shortest-path routing. The scheduling policies included combinations of FIFO, SP, WFQ, and DRR with a fixed queue size of 32 packets, and the traffic profile consisted of Poisson arrivals, binomial packet sizes, and randomly assigned Type of Service (ToS) per path from values 0, 1, and 2.

\subsection{Scalability Dataset}
To evaluate the model's scalability and its ability to generalize to larger networks, we used datasets with varying network sizes. The training topologies included randomly generated networks with sizes ranging from 25 to 50 nodes, while the testing topologies ranged from 50 to 300 nodes. All networks used FIFO scheduling with a queue size of 32 packets. The traffic profile included Poisson time distributions and binomial packet sizes between 300 and 1,700 bits. For each sample, a maximum average lambda was selected between 400 and 2,000 bps (representing traffic intensity), with 400 being the lowest congested network (0\% avg. packet loss) and 2,000 a highly congested network (3\% avg. packet loss). The bandwidth per path was determined by multiplying this lambda with a uniform random value between 0.1 and 1.

\subsection{Real Traffic and Traffic Models Datasets}
To test the model under different traffic conditions, we used datasets with varied time and size distributions. The Traffic Model dataset featured time distributions such as Constant Bit Rate (CBR), On-Off traffic, autocorrelated exponentials, modulated exponentials, and mixed traffic profiles, utilizing NSFNET and GEANT2 topologies for training and GBN for testing, all with binomial packet size distributions.

The Real Traffic dataset employed traffic profiles derived from actual network traces from the MAWI repository (Sample point 2022/09) \cite{31}, with training conducted on the GEANT topology and testing on ABILENE, GERMANY50, and NOBEL topologies. The traffic matrices were based on data from the SNDlib library \cite{32}, scaled for computational feasibility. We used a distribution of source-destination flows from a real internet service provider \cite{33} to map flows to different ToS classes. Scheduling policies included combinations of FIFO, SP, WFQ, and DRR, with queue sizes randomly selected per node from 8,000 to 64,000 bits. For this experiment, we leveraged knowledge from previous versions and used a previous checkpoint, fine-tuning it using 200 samples of the GEANT topology.

\section{Network Tasks}

Our implementation of RouteNet-Fermi model with its three model variants (RNN, GRU, and LSTM) is evaluated on three key network performance prediction tasks. The primary task involves predicting the end-to-end delay for network flows. This consists of two components: queuing delay which is time packets spend waiting in router queues, and transmission delay which is time required to transmit packets across links. The total delay is the sum of these components, and this task is challenging as it requires understanding both network congestion patterns and the impact of different traffic models. The next task is jitter prediction, which involves estimating the variance in packet delay. The model predicts jitter by analyzing delay variations across time and considering traffic burstiness and queue occupancy fluctuations, accounting for different scheduling policies' impact on delay variation. The jitter prediction task requires additional readout functions and is particularly sensitive to the RNN cell's ability to capture temporal dependencies. The final task is packet loss prediction which focuses on predicting the probability of packet loss in network flows. The main aspects include estimating buffer overflow probabilities and considering queue occupancy patterns, accounting for scheduling policy effects on buffer management. The three prediction tasks are inherently related and influence each other. The Delay-Jitter relationship shows that higher average delays correlate with increased jitter and queue occupancy affects both metrics, but traffic burstiness mainly impacts both delay and jitter. The Delay-Loss relationship shows high delays can indicate imminent packet loss, and buffer overflow causes both increased delays and losses, but queue management policies affect both metrics. Finally the Loss-Jitter relationship shows packet losses can cause sudden jitter spikes, but both are sensitive to traffic bursts, and buffer dynamics also influence both metrics. These tasks together provide a comprehensive evaluation of the model's ability to capture different aspects of network performance, with each RNN variant potentially showing different strengths across the tasks.

\section{Deep Learning Methods}

\begin{figure}[h]
    \centering
    \includegraphics[width=\linewidth]{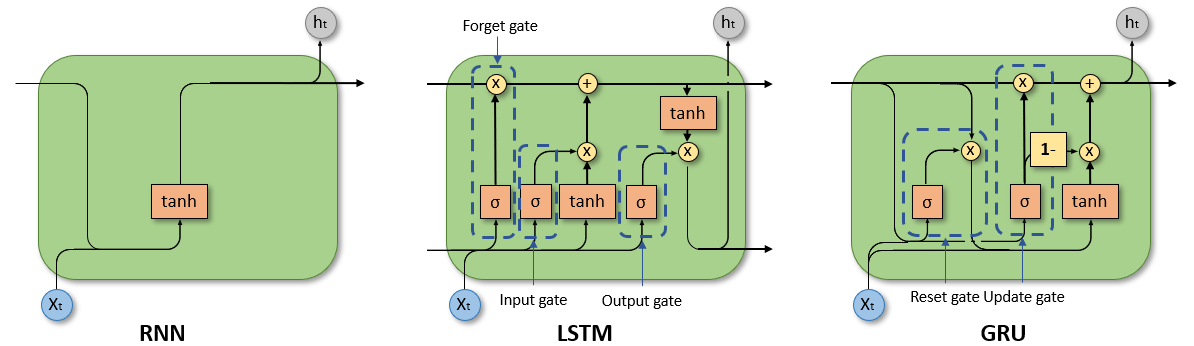}
    \caption{Comparison of neural network cell architectures. (a) RNN cell showing simple structure with a single hidden state. (b) LSTM cell with input, forget, output gates and memory cell state. (c) GRU cell with reset and update gates.}
    \label{fig:model_arch}
\end{figure}

RouteNet-Fermi employs a three-stage message-passing architecture to model the complex interactions between network flows, queues, and links. The model iteratively updates hidden state representations of these components to capture their evolving relationships. At the core of this architecture are recurrent neural networks, which process sequential information about network states. Our implementation extends the original model by introducing three different recurrent cell variants: RNN, GRU, and LSTM. Figure \ref{fig:model_arch} shows the architecture of the three cells. We integrate these cells into both the Flow-Level RNN and Link-Level RNN components of RouteNet-Fermi. 

The RNN cell is the most fundamental recurrent architecture and it processes input sequences using a single transformation layer. For timestep $t$, given an input $x_t$ and previous hidden state $h_{t-1}$, the simple RNN computes:

$$h_t = tanh(W_h \cdot [h_{t-1}, x_t] + b_h)$$

Where $W_h$ represents the weight matrix for the hidden state, $[h_{t-1}, x_t]$ denotes the concatenation of previous hidden state and current input, and $b_h$ is the bias term tanh is the hyperbolic tangent activation function. For input dimension $d$ and hidden state dimension $h$ and  sequence of length $T$, the space complexity is $O(dh + h^2)$ and time complexity is $O(T(dh + h^2))$. RNNs are straightforward and computationally efficient but struggle with long-term dependencies due to vanishing gradient problem.

GRU cells which is used in the original RouteNet-Fermi implementation introduces gating mechanisms to better control information flow. The GRU computation involves the following equations:

$$z_t = \sigma(W_z \cdot [h_{t-1}, x_t] + b_z)  \texttt{[Update gate]}$$
$$r_t = \sigma(W_r \cdot [h_{t-1}, x_t] + b_r)  \texttt{[Reset gate]}$$
$$h_t' = tanh(W \cdot [r_t\odot h_{t-1}, x_t] + b) \texttt{[Candidate hidden state]}$$
$$h_t = (1 - z_t) \odot h_{t-1} + z_t \odot h_t'   \texttt{[Final hidden state]}$$

Where $\sigma$ represents the sigmoid activation function, $\odot$ denotes element-wise multiplication, $W_z, W_r, W$ are weight matrices, and $b_z, b_r, b$ are bias terms. For input dimension $d$ and hidden state dimension $h$ and  sequence of length $T$, the space complexity is $O(3(dh + h^2))$ and time complexity is $O(3T(dh + h^2))$. GRU cells have better gradient flow compared to RNN and is effective for medium-range dependencies.

LSTM cells introduce an additional memory cell state and more sophisticated gating mechanisms. The LSTM computation follows:

$$f_t = \sigma(W_f \cdot (h_{t-1}, x_t) + b_f) \texttt{[Forget gate]}$$
$$i_t = \sigma(W_i \cdot [h_{t-1}, x_t] + b_i) \texttt{[Input gate]}$$
$$o_t = \sigma(W_o \cdot [h_{t-1}, x_t] + b_o) \texttt{[Output gate]}$$
$$c_t' = tanh(W_c \cdot [h_{t-1}, x_t] + b_c)  \texttt{[Candidate cell state]}$$
$$c_t = f_t \odot c_{t-1} + i_t \odot c_t'  \texttt{[Cell state update]}$$
$$h_t = o_t \odot tanh(c_t)  \texttt{[Hidden state update]}$$

Where $c_t$ represents the cell state, $W_f, W_i, W_o, W_c$ are weight matrices, and $b_f, b_i, b_o, b_c$ are bias terms. For input dimension $d$ and hidden state dimension $h$ and  sequence of length $T$, the space complexity is $O(4(dh + h^2))$ and time complexity is $O(4T(dh + h^2))$. LSTM cells have the most sophisticated control over information flow and are suitable for complex temporal patterns.

\subsection{Training Configuration}

All the deep learning models were implemented using Tensorflow Python library and trained on the CPU. The pipeline how we approached this analysis was to download the given datasets used in the paper and forking their original repository. Then we installed all the required dependencies and resolved any deprecation errors. Finally we moved on to running the available GRU models for each dataset, trying to replicate the results of the paper. Once we gathered all the reproduction results we moved on to implement the LSTM and RNN models to compare the results against the paper's methods. For fair and consistent comparison we maintained the same parameters across all cell variants with hidden states set to 32, batch size set to 2,000 samples, training epochs set to task specific numbers, optimizer set to Adam with learning rate of 0.001. The loss functions remain specific to each prediction task: delay and jitter with Mean Absolute Percentage Error (MAPE), and packet loss with Mean Absolute Error (MAE).

\section{Results}

Our evaluation compares three RNN variants (RNN, LSTM, and GRU) across multiple network scenarios against the original RouteNet-Fermi GRU implementation. The results demonstrate distinct performance characteristics for each architecture across different network tasks and topologies. 

\subsection{Fat Tree}

\begin{figure}[H]
    \centering
    \begin{subfigure}[b]{0.32\linewidth}
   \centering
   \includegraphics[width=\linewidth]{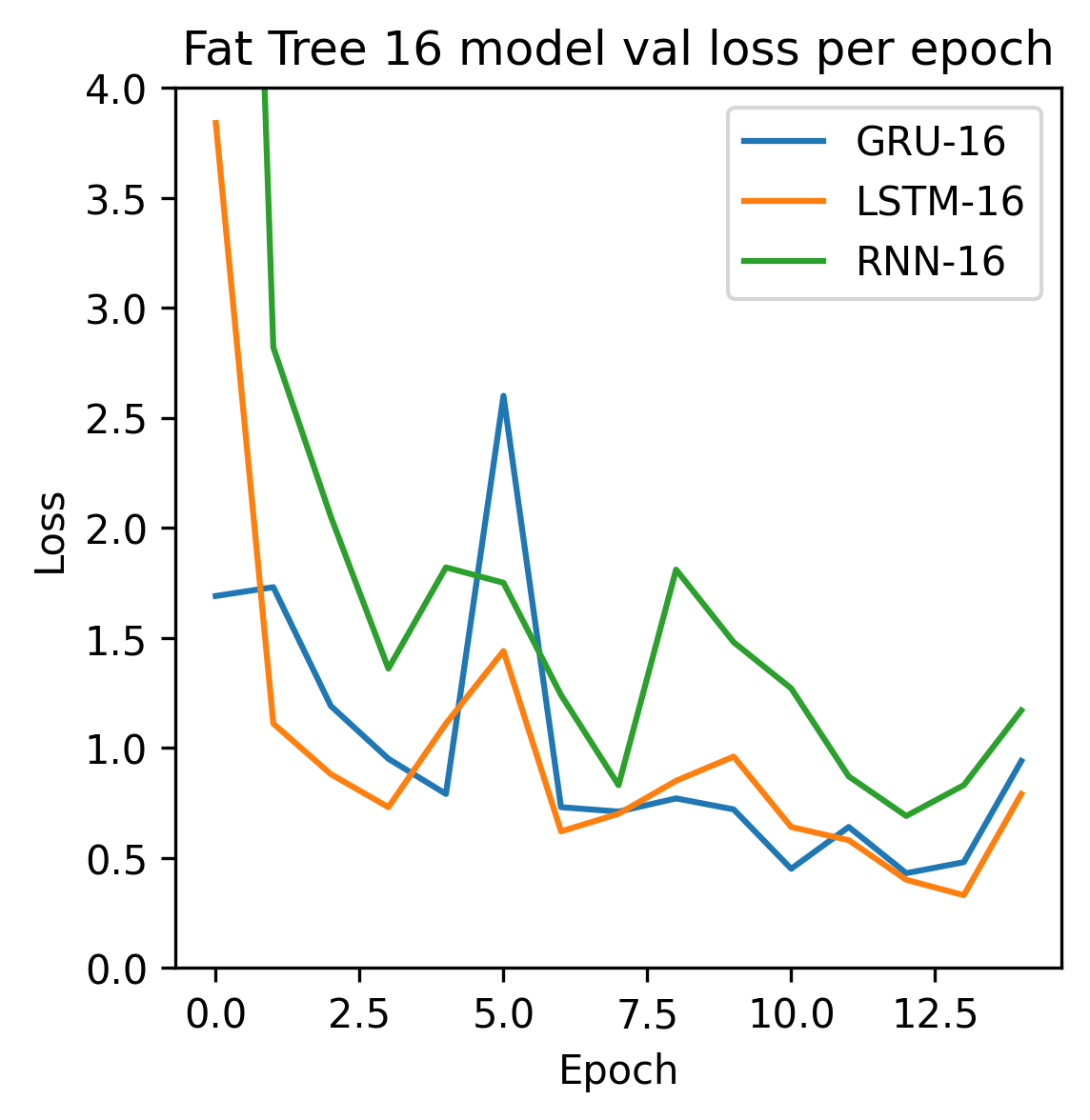}
   \caption{FatTree16}
   \label{fig:fat16}
    \end{subfigure}
    \hfill
    \begin{subfigure}[b]{0.32\linewidth}
   \centering
   \includegraphics[width=\linewidth]{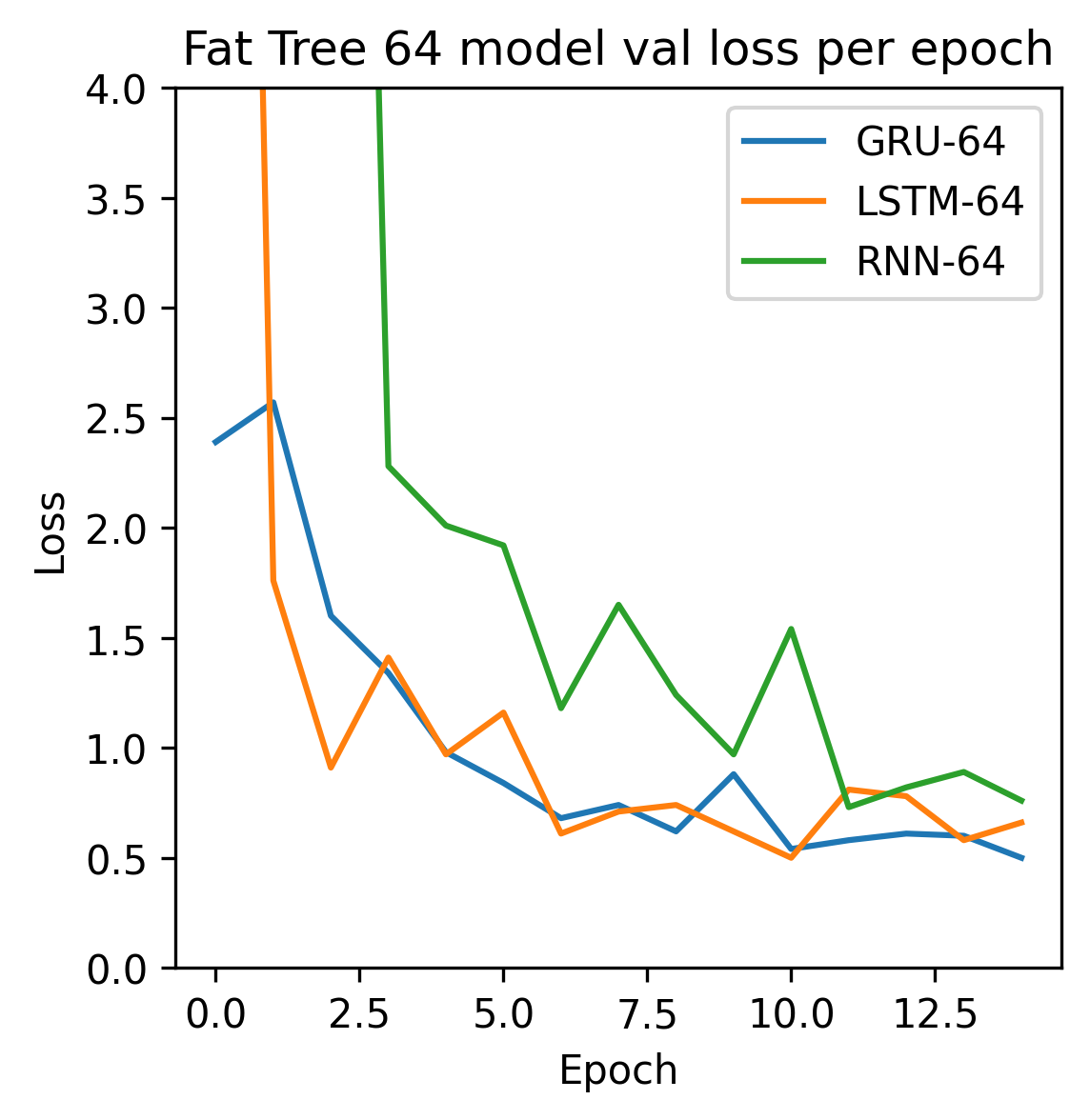}
   \caption{FatTree64}
   \label{fig:fat64}
    \end{subfigure}
    \hfill
    \begin{subfigure}[b]{0.32\linewidth}
   \centering
   \includegraphics[width=\linewidth]{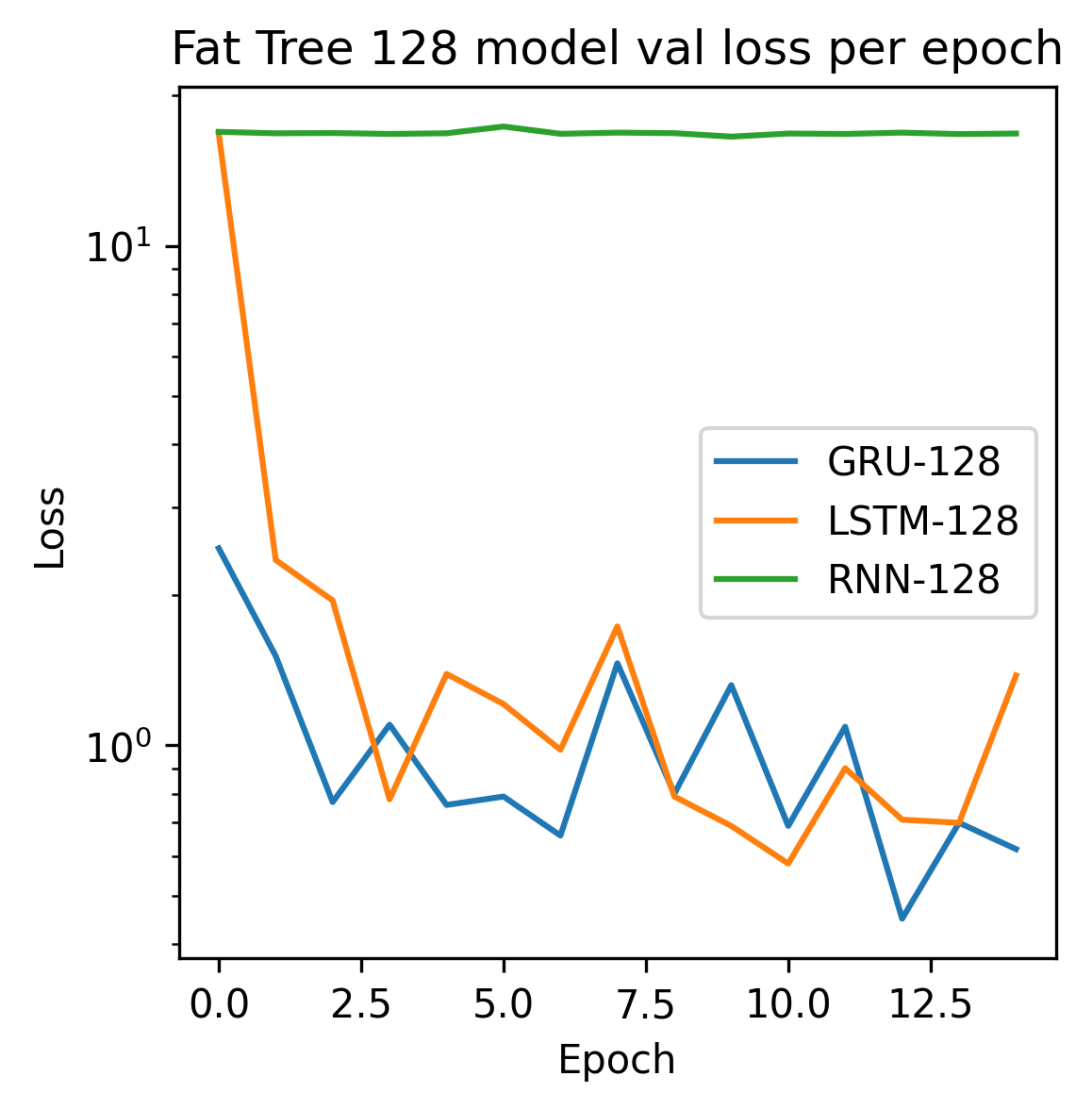}
   \caption{FatTree128}
   \label{fig:fat128}
    \end{subfigure}
    \caption{Comparison of model validation loss per epoch for Fat Tree topology for delay prediction }
    \label{fig:fattree}
\end{figure}

In Fat Tree topology evaluations shown in table \ref{tab:fattree}, all three models showed strong performance, with LSTM cells consistently achieving the lowest MAPE. For FatTree16: the LSTM model achieved 0.33\% MAPE vs RNN's 0.69\% and GRU's 0.43\%. For FatTree64: the original paper GRU showed stronger performance with 0.44\% MAPE. For FatTree128: the LSTM and GRU models both showed great stability (0.58\% and 0.45\% respectively) while RNN's performance degraded significantly to 16.53\% MAPE. The validation loss curves in figure \ref{fig:fattree} show faster convergence for LSTM and GRU models compared to RNN, particularly in larger topologies.

\begin{table}[H]
\caption{Comparison of model metrics on test dataset for Fat Tree topology for delay prediction}
\resizebox{0.49\textwidth}{!}{
\begin{tabular}{l|ccc|c}
\cmidrule(r){1-5}
Topology & RNN Model & LSTM Model & GRU Model & Paper GRU \\ 
&MAPE&MAPE&MAPE&MAPE \\
\cmidrule(r){1-5}
FatTree16& 0.69\%  & \textbf{0.33\%}  & 0.43\%& 0.37\% \\
FatTree64& 0.73\%  & 0.50\%& 0.50\%& \textbf{0.44\%}   \\
FatTree128& 16.53\%& 0.58\%& \textbf{0.45\%}& 0.67\%   \\ 
\cmidrule(r){1-5}
\end{tabular}
\label{tab:fattree}
}
\end{table}

\subsection{Real Traffic and Scalability}

\begin{figure}[H]
    \centering
    \begin{subfigure}[b]{0.24\linewidth}
    \centering
    \includegraphics[width=\linewidth]{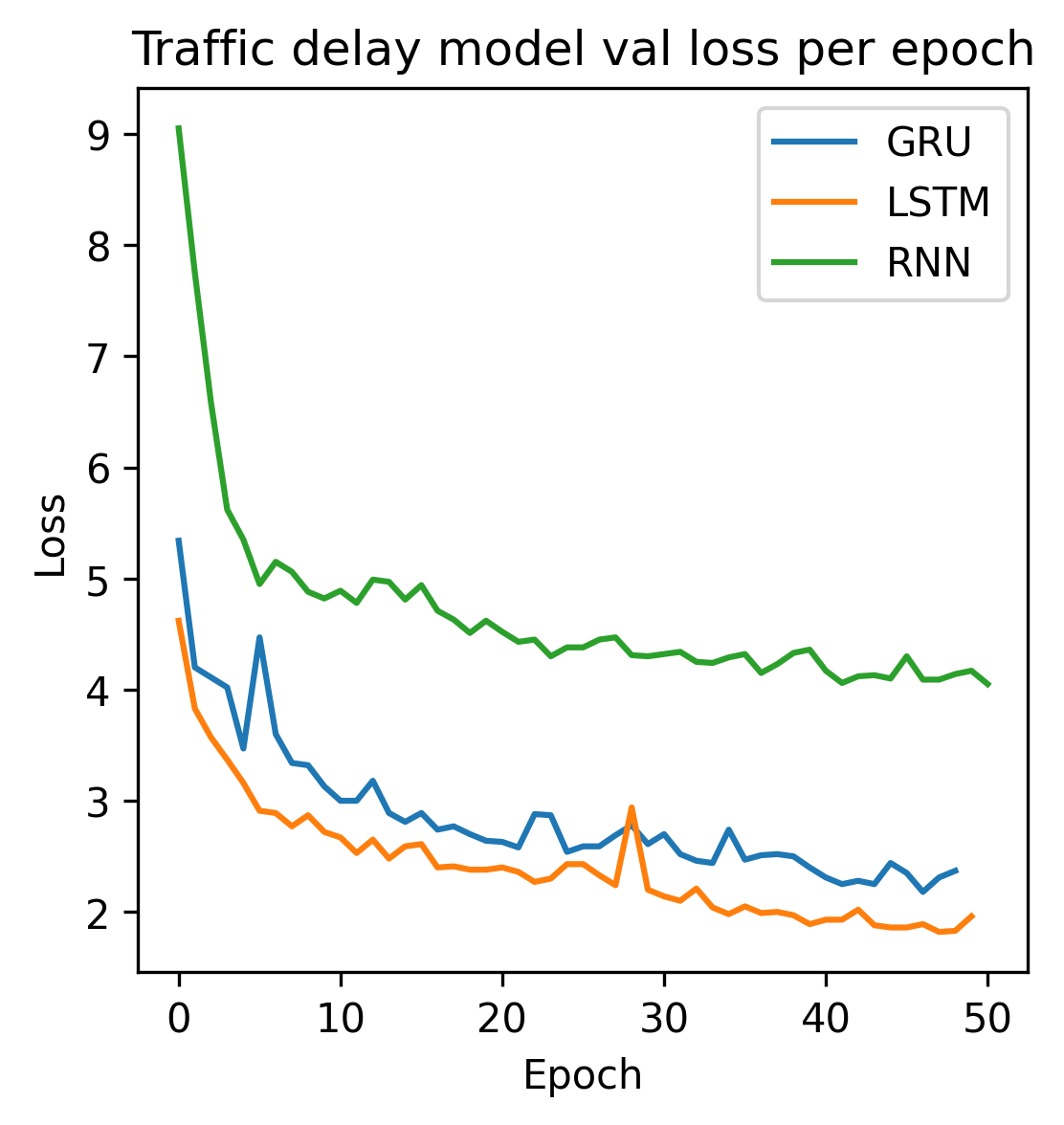}
    \caption{Delay}
    \label{fig:realDelay}
    \end{subfigure}
    \begin{subfigure}[b]{0.24\linewidth}
    \centering
    \includegraphics[width=\linewidth]{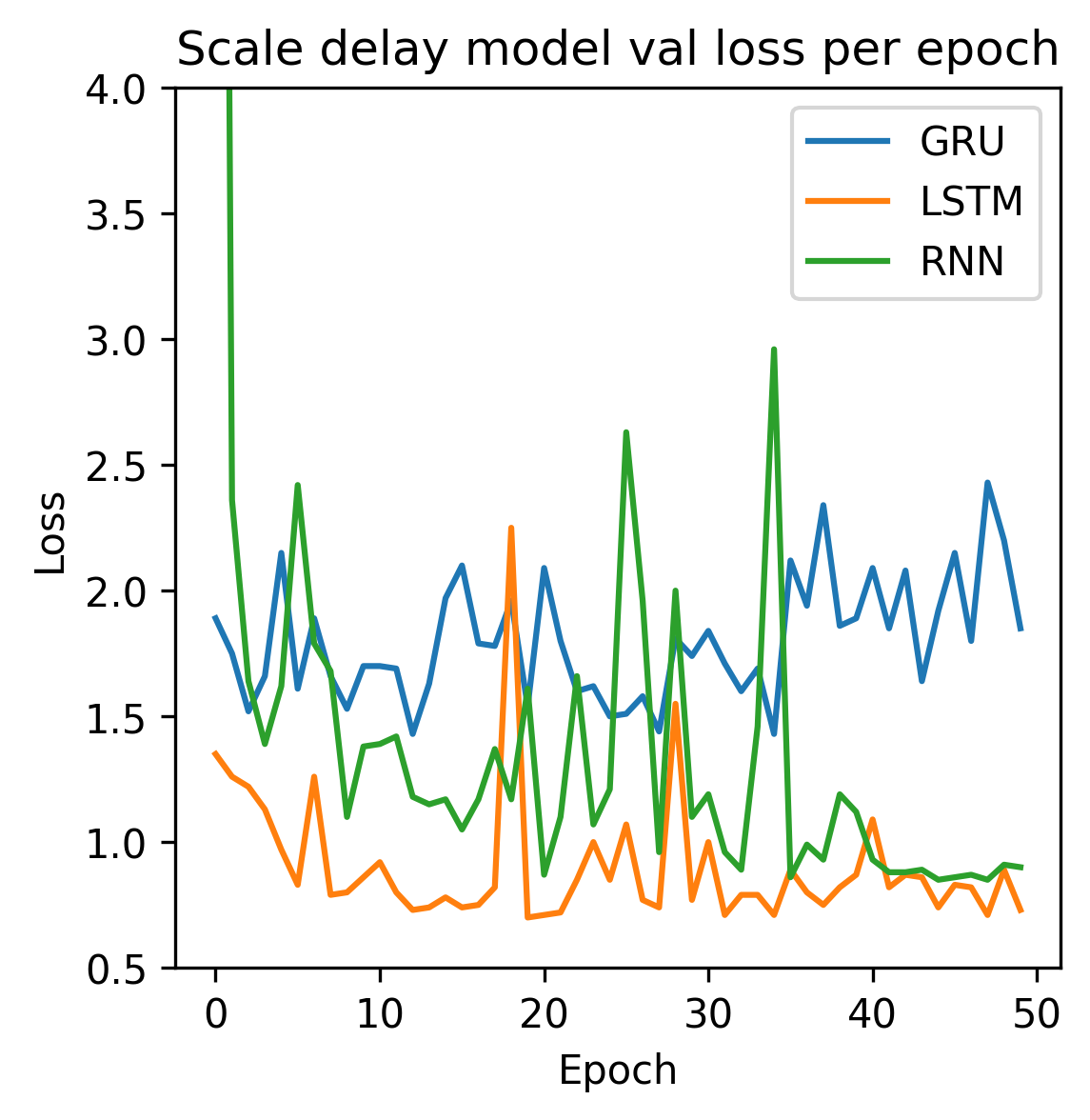}
    \caption{Delay}
    \label{fig:scaleDelay}
    \end{subfigure}
    \begin{subfigure}[b]{0.24\linewidth}
    \centering
    \includegraphics[width=\linewidth]{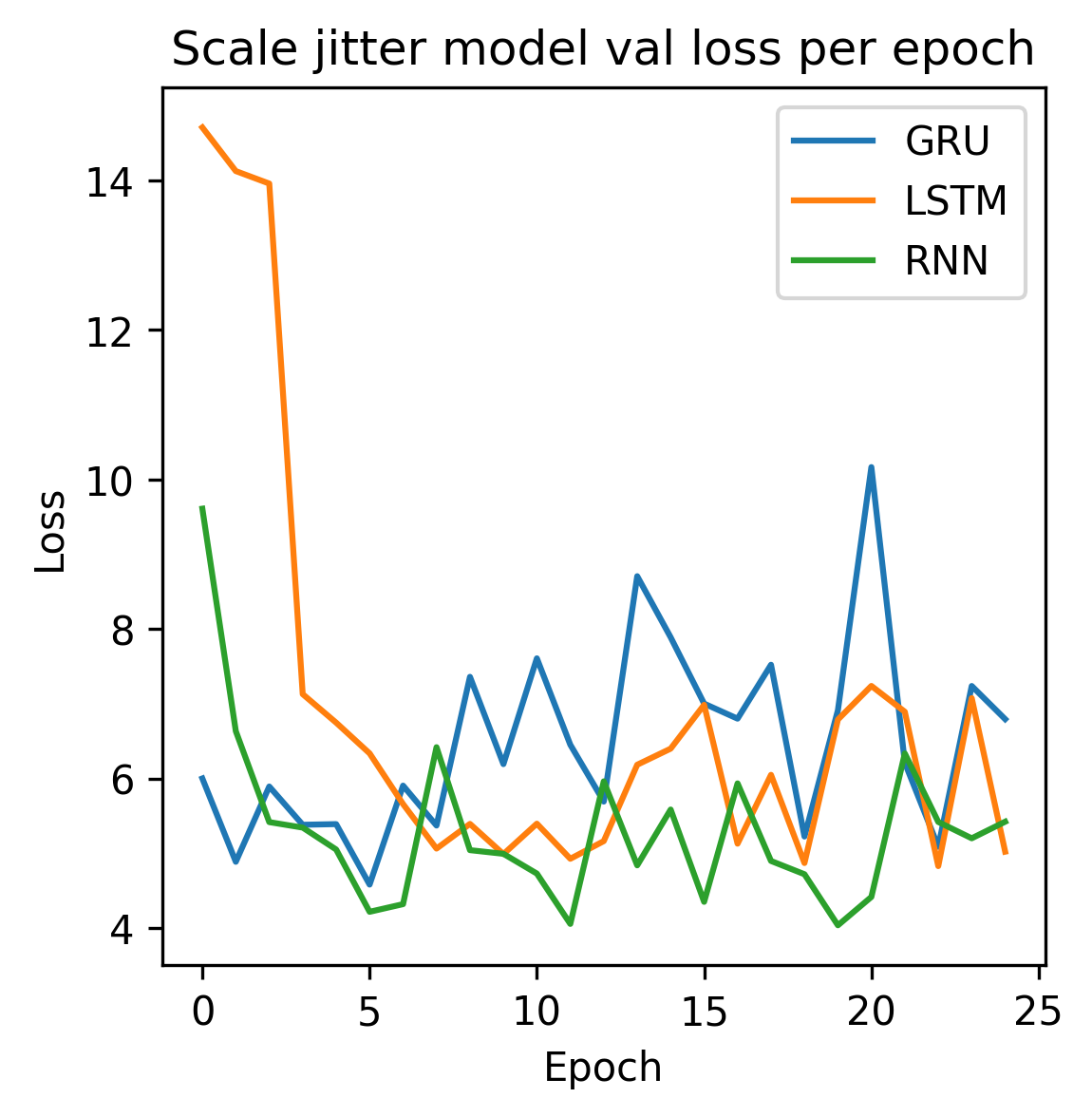}
    \caption{Jitter}
    \label{fig:scaleJitter}
    \end{subfigure}
    \begin{subfigure}[b]{0.24\linewidth}
    \centering
    \includegraphics[width=\linewidth]{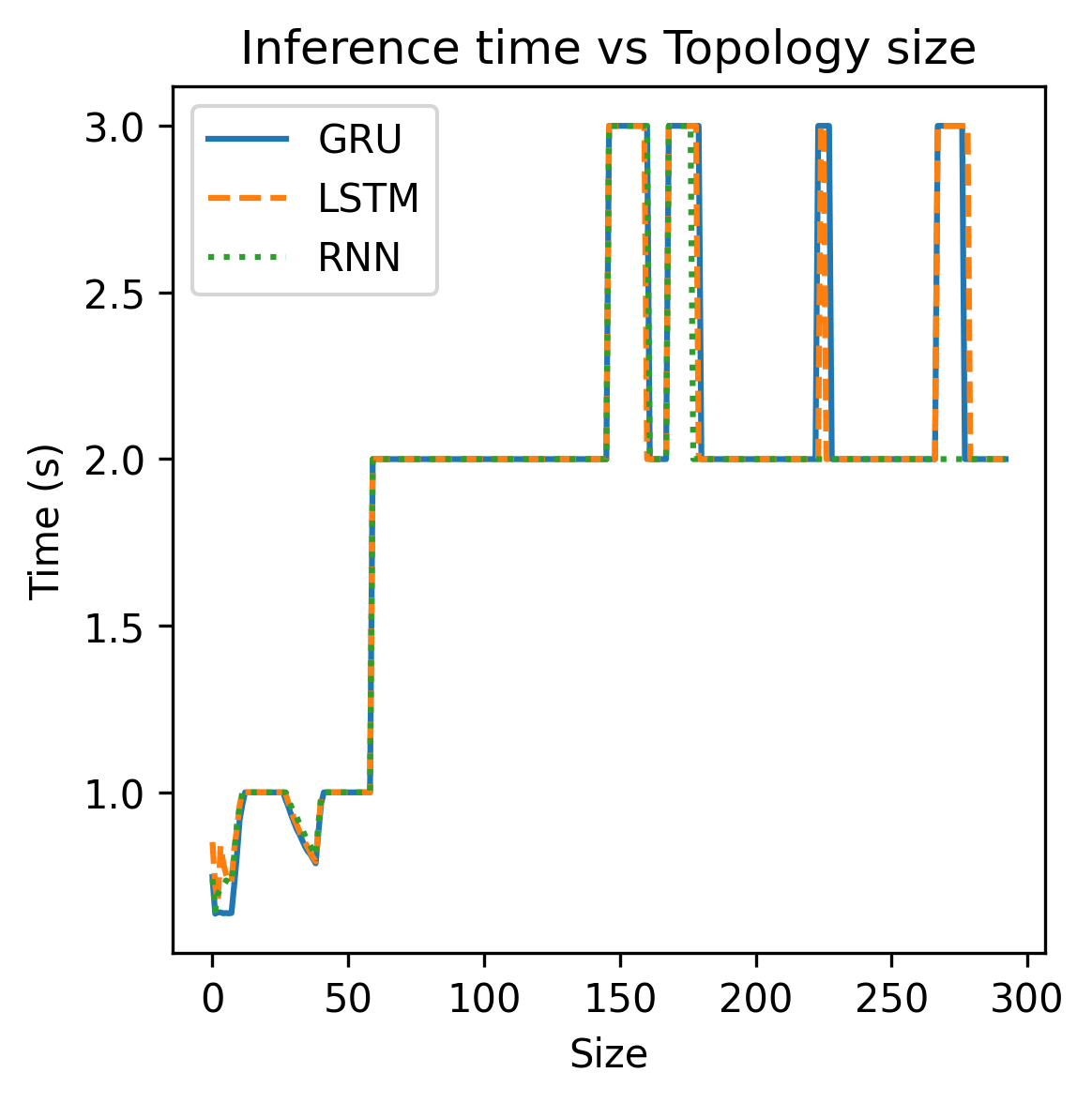}
    \caption{Inference}
    \label{fig:scaletime}
    \end{subfigure}
    \caption{Comparison of model validation loss per epoch for Real Traffic topology and Scalability}
    \label{fig:scalability}
\end{figure}

In real traffic scenarios shown in \ref{tab:traffic}, the LSTM implementation achieved a MAPE of 1.82\%, outperforming both RNN (4.05\%) and GRU (2.18\%) variants, as well as the original paper's reported 5.67\%. This suggests that LSTM's sophisticated memory mechanisms better capture the temporal patterns in real network traffic. The scalability evaluation shown in \ref{tab:scale} shows interesting trade-offs between model architectures where for delay prediction LSTM showed the best performance with 0.70\% MAPE, while for jitter prediction RNN achieved 3.96\% MAPE, slightly better than LSTM's 4.83\%. The model validation loss curves in figure \ref{fig:realDelay} show that the LSTM and GRU models converge better and faster than the RNN model. While in figures \ref{fig:scaleDelay} and \ref{fig:scaleJitter} the models do not seem to converge and fluctuate around the same point. The figure \ref{fig:scaletime} shows how the inference time increases per model based on increasing topology size from nodes 10 to 300.

\begin{table}[H]
\caption{Comparison of model metrics on test dataset for Real Traffic topology for delay prediction}
\resizebox{0.49\textwidth}{!}{
\begin{tabular}{c|ccc|c}
\cmidrule(r){1-5}
& RNN Model & LSTM Model      & GRU Model & Paper GRU \\
\cmidrule(r){1-5}
MAPE & 4.05\%    & \textbf{1.82\%} & 2.18\%    & 5.67\% \\
\cmidrule(r){1-5}
\end{tabular}
\label{tab:traffic}
}
\end{table}

\begin{table}[h]
\caption{Comparison of model metrics on test dataset for Scalability topology for delay and jitter prediction}
\resizebox{0.49\textwidth}{!}{
\begin{tabular}{c|ccc|c}
\cmidrule(r){1-5}
& RNN Model & LSTM Model & GRU Model & Paper GRU \\
\cmidrule(r){1-5}
Delay MAPE & 0.85\% & \textbf{0.70\%} & 1.43\% & 1.08\%\\
Jitter MAPE & \textbf{3.96\%} & 4.83\% & 4.58\% & 4.54\%\\
\cmidrule(r){1-5}
\end{tabular}
\label{tab:scale}
}
\end{table}

\subsection{Scheduling}

\begin{figure}[H]
    \centering
    \begin{subfigure}[b]{0.32\linewidth}
   \centering
   \includegraphics[width=\linewidth]{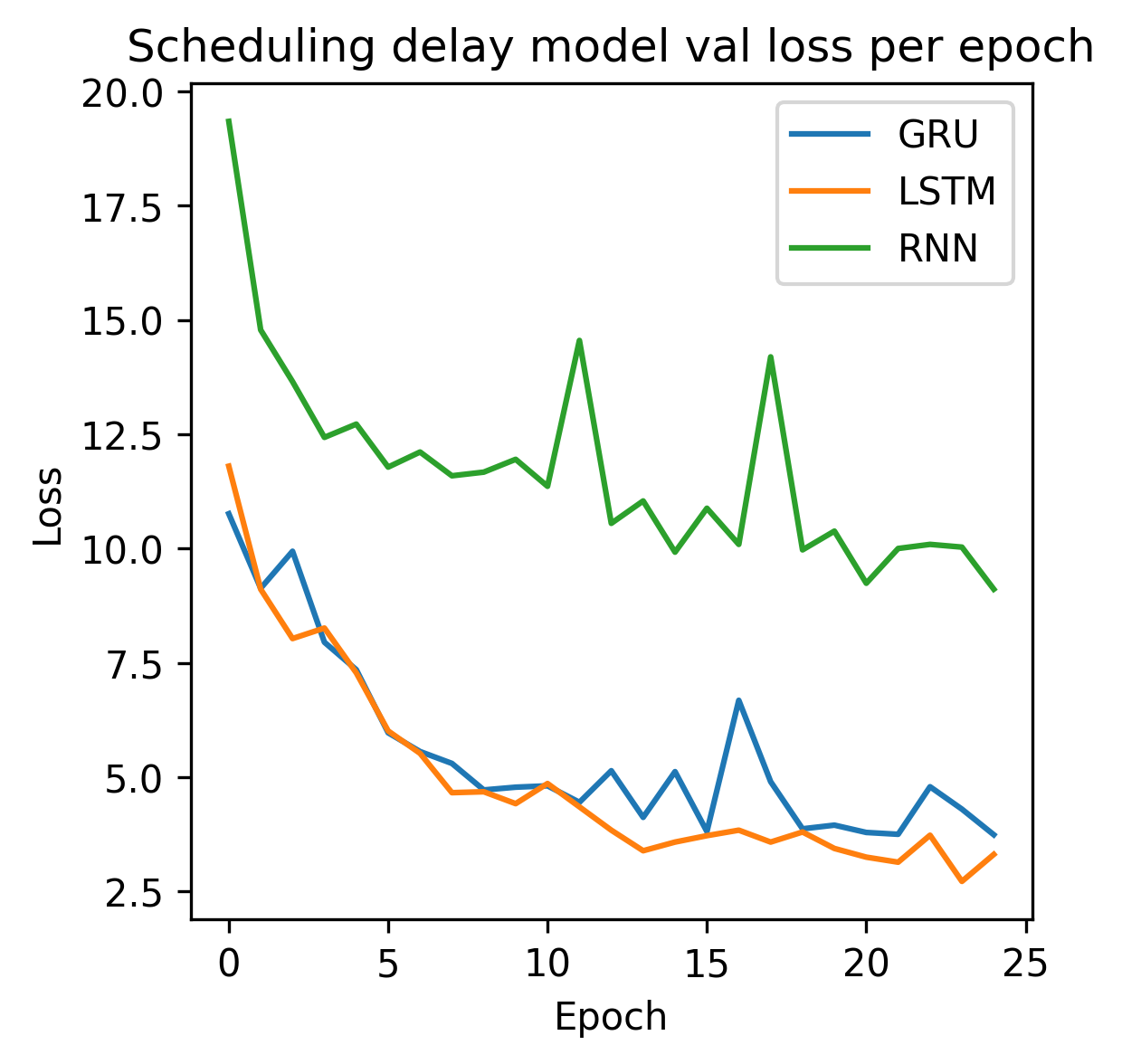}
   \caption{Delay}
   \label{fig:scheddelay}
    \end{subfigure}
    \hfill
    \begin{subfigure}[b]{0.31\linewidth}
   \centering
   \includegraphics[width=\linewidth]{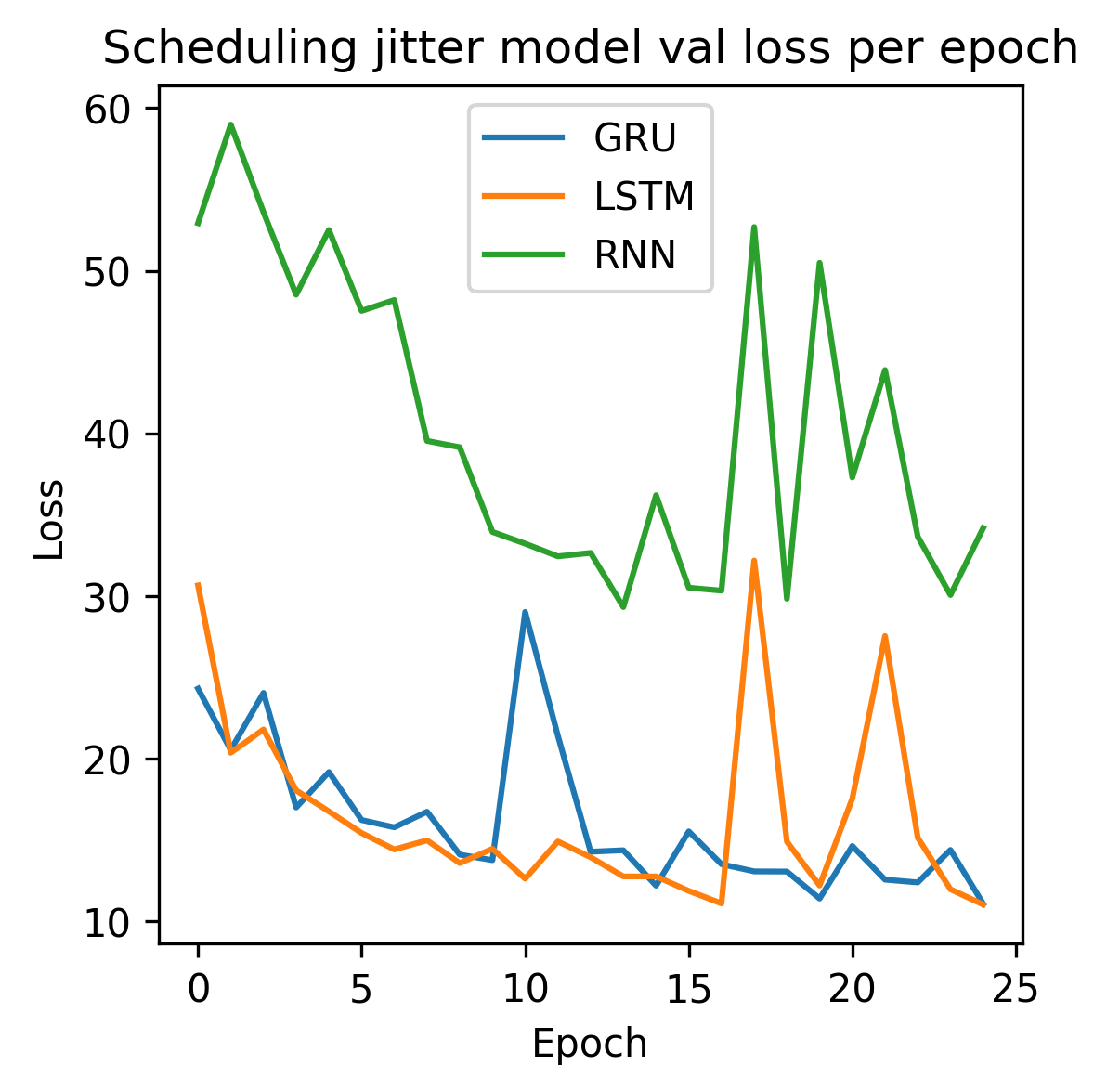}
   \caption{Jitter}
   \label{fig:scheddelay1}
    \end{subfigure}
    \hfill
    \begin{subfigure}[b]{0.33\linewidth}
   \centering
   \includegraphics[width=\linewidth]{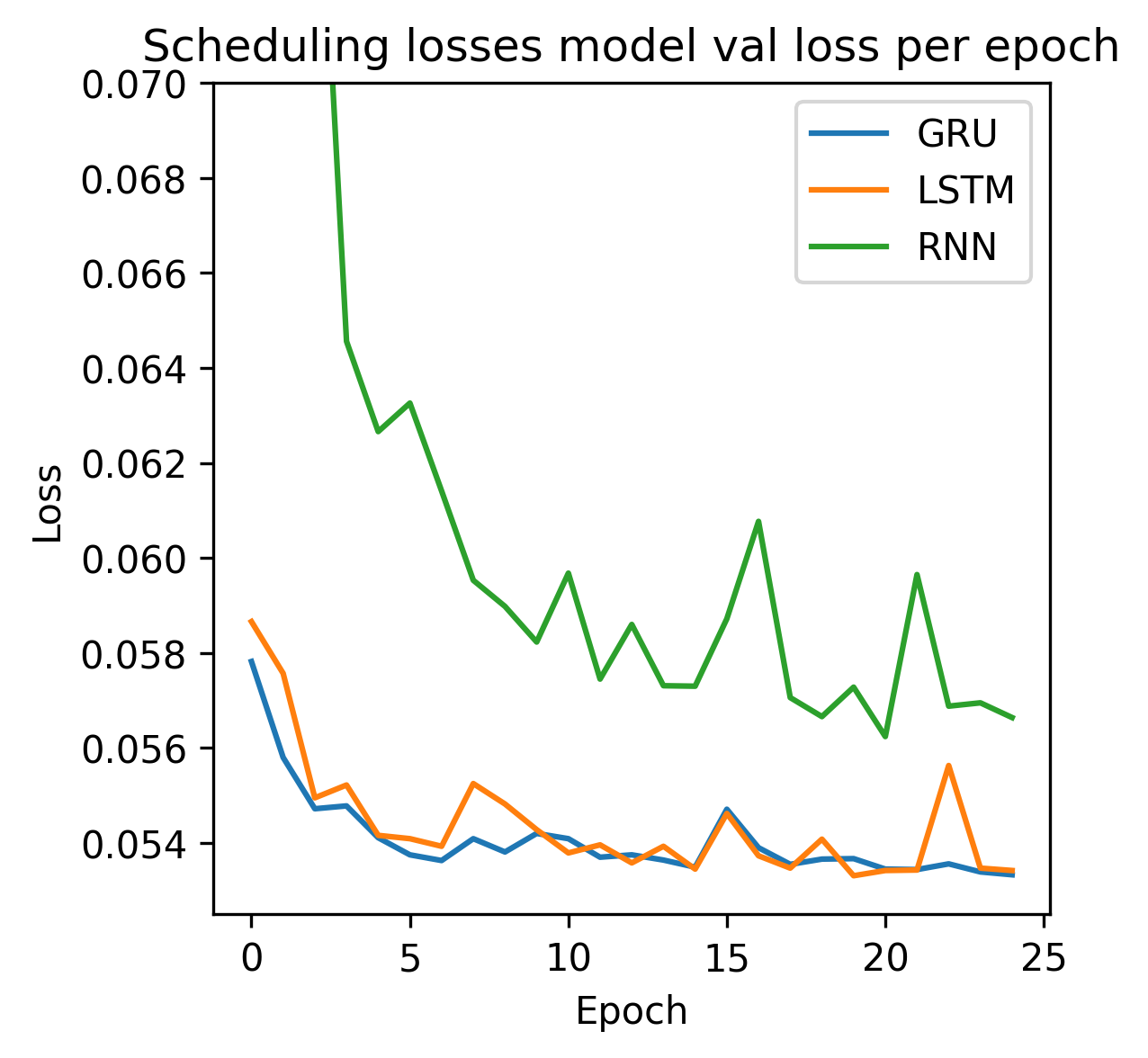}
   \caption{Losses}
   \label{fig:scheddelay2}
    \end{subfigure}
    \caption{Comparison of delay model validation loss per epoch with different Scheduling policies}
    \label{fig:scheduling}
\end{figure}

For networks with different scheduling policies in table \ref{tab:schedule} the delay prediction task showed that LSTM significantly outperformed other variants with 2.96\% MAPE and for jitter prediction task LSTM achieved 16.73\% MAPE. For the packet loss prediction task the original paper GRU showed better performance with MAE of 0.001978. The validation loss curves in figure \ref{fig:scheduling} demonstrate that LSTM and GRU models show more stable convergence patterns compared to RNN across all three metrics.

\begin{table}[H]
\caption{Comparison of model metrics on test dataset with Scheduling policies for delay, jitter and losses prediction}
\resizebox{0.49\textwidth}{!}{
\begin{tabular}{c|ccc|c}
\cmidrule(r){1-5}
& RNN Model & LSTM Model & GRU Model & Paper GRU \\
\cmidrule(r){1-5}
Delay MAPE & 9.80\% & \textbf{2.96\%} & 3.82\% & 3.35\% \\
Jitter MAPE & 29.70\% & \textbf{16.73\%} & 17.01\% & 17.21\%\\
Losses MAE & 0.007464 & 0.002218 & 0.002063 & \textbf{0.001978}\\
\cmidrule(r){1-5}
\end{tabular}
\label{tab:schedule}
}
\end{table}

\subsection{Traffic Models}

\begin{figure}[ht]
    \centering
    \begin{subfigure}[b]{0.32\linewidth}
   \centering
   \includegraphics[width=\linewidth]{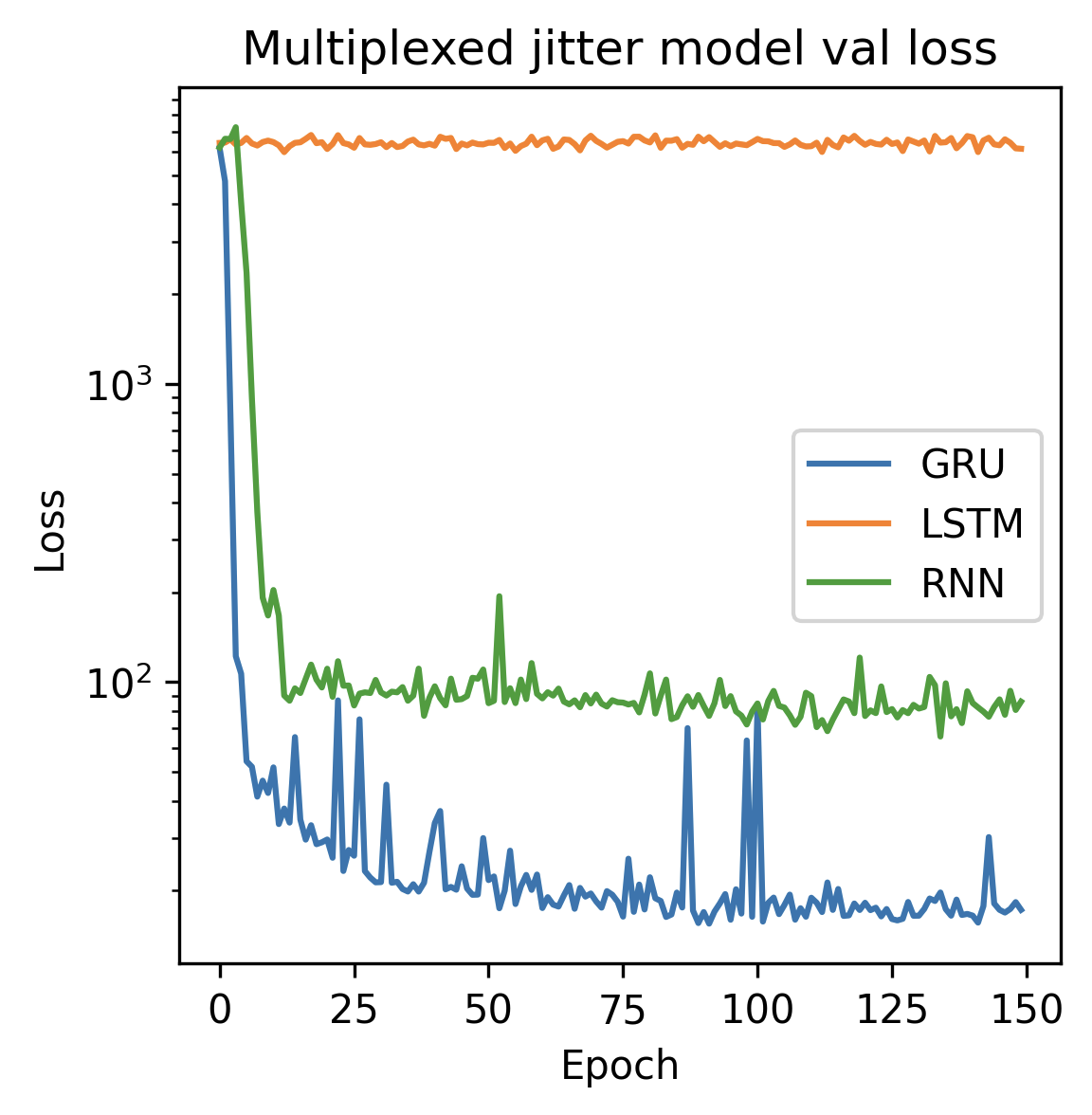}
   \caption{multi-jitter}
   \label{fig:multijitter}
    \end{subfigure}
    \hfill
    \begin{subfigure}[b]{0.32\linewidth}
   \centering
   \includegraphics[width=\linewidth]{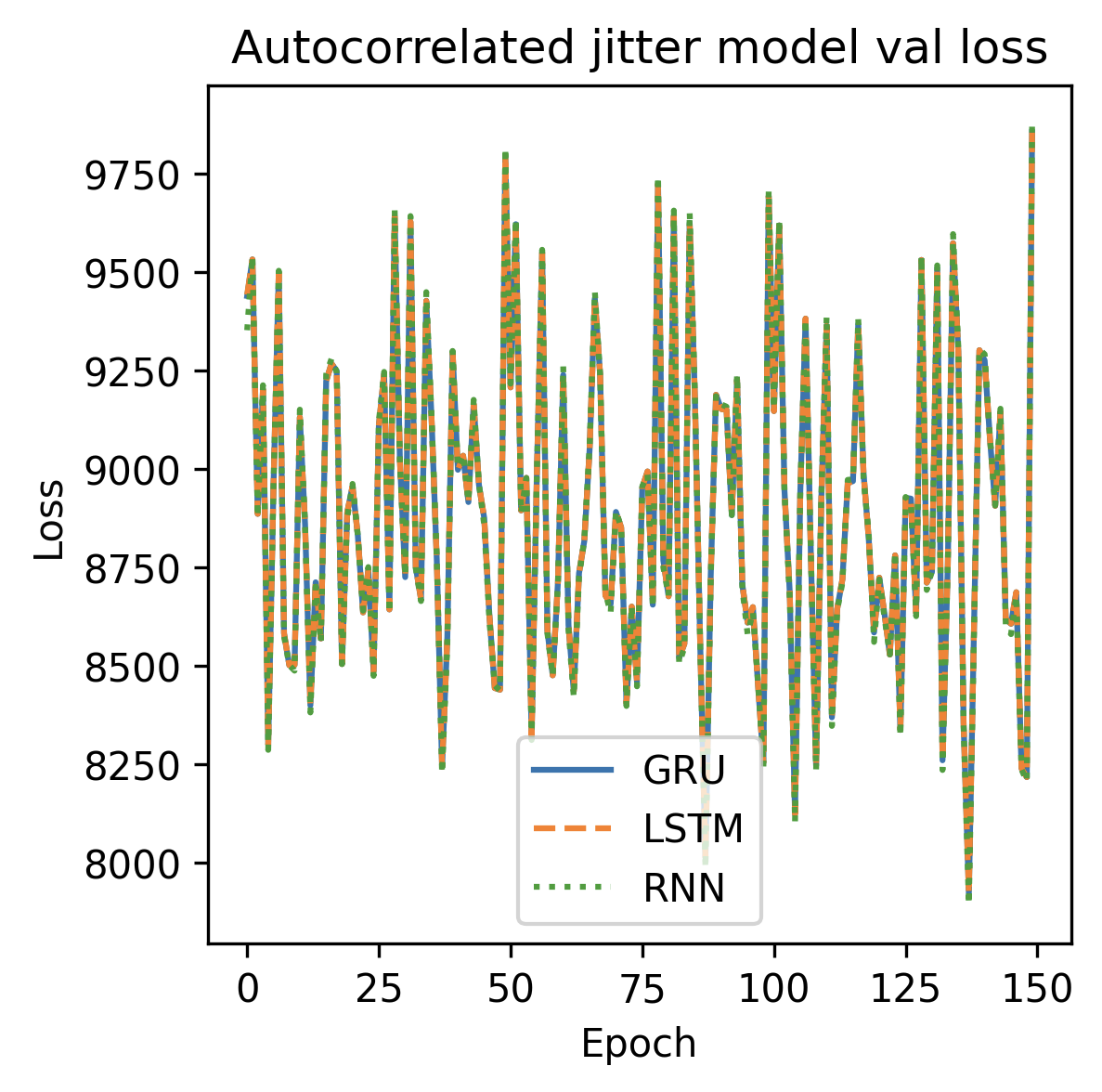}
   \caption{auto-jitter}
   \label{fig:autojitter}
    \end{subfigure}
    \hfill
    \begin{subfigure}[b]{0.32\linewidth}
   \centering
   \includegraphics[width=\linewidth]{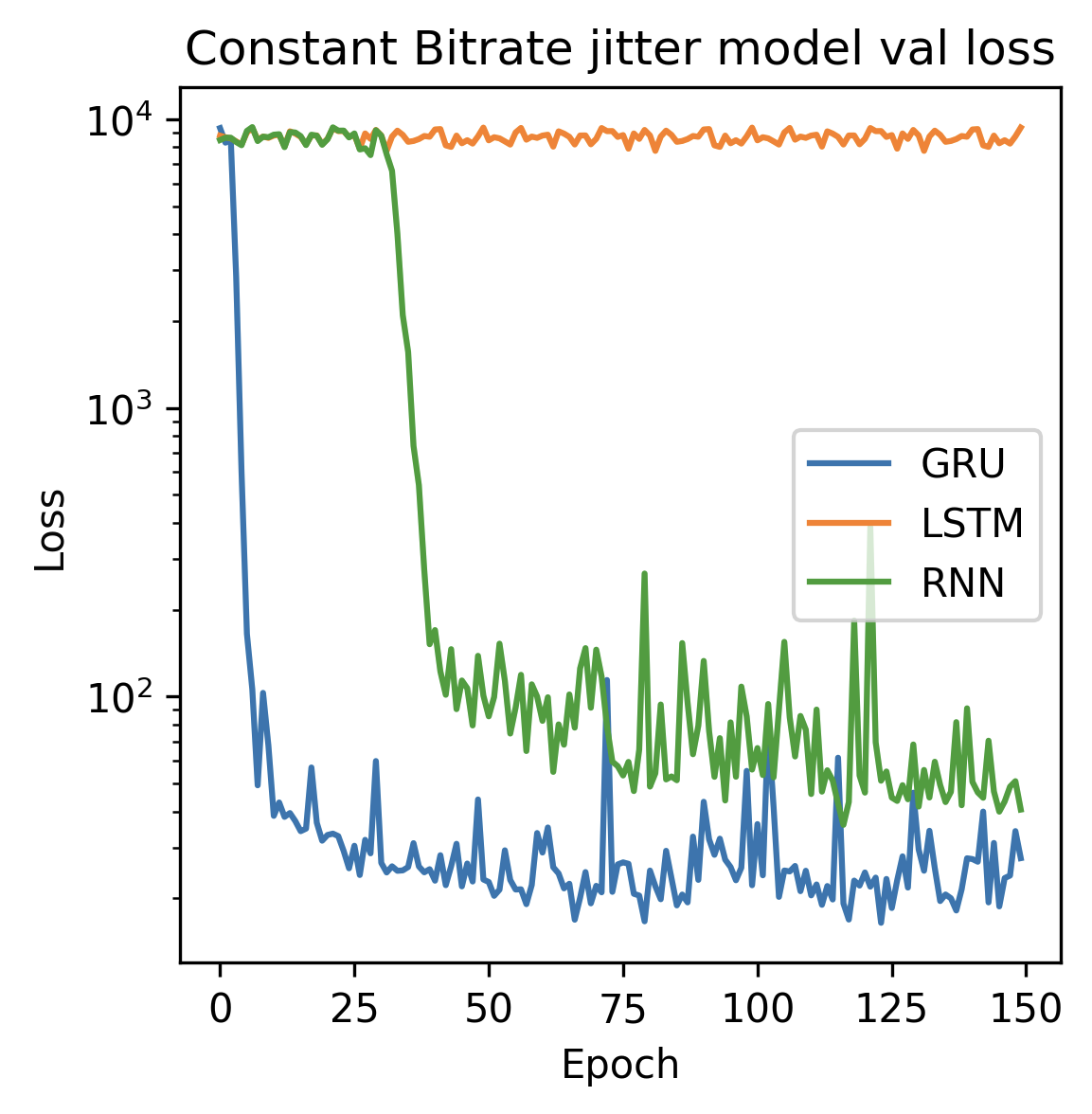}
   \caption{bit-jitter}
   \label{fig:bitjitter}
    \end{subfigure}
    \hfill
    \begin{subfigure}[b]{0.32\linewidth}
   \centering
   \includegraphics[width=\linewidth]{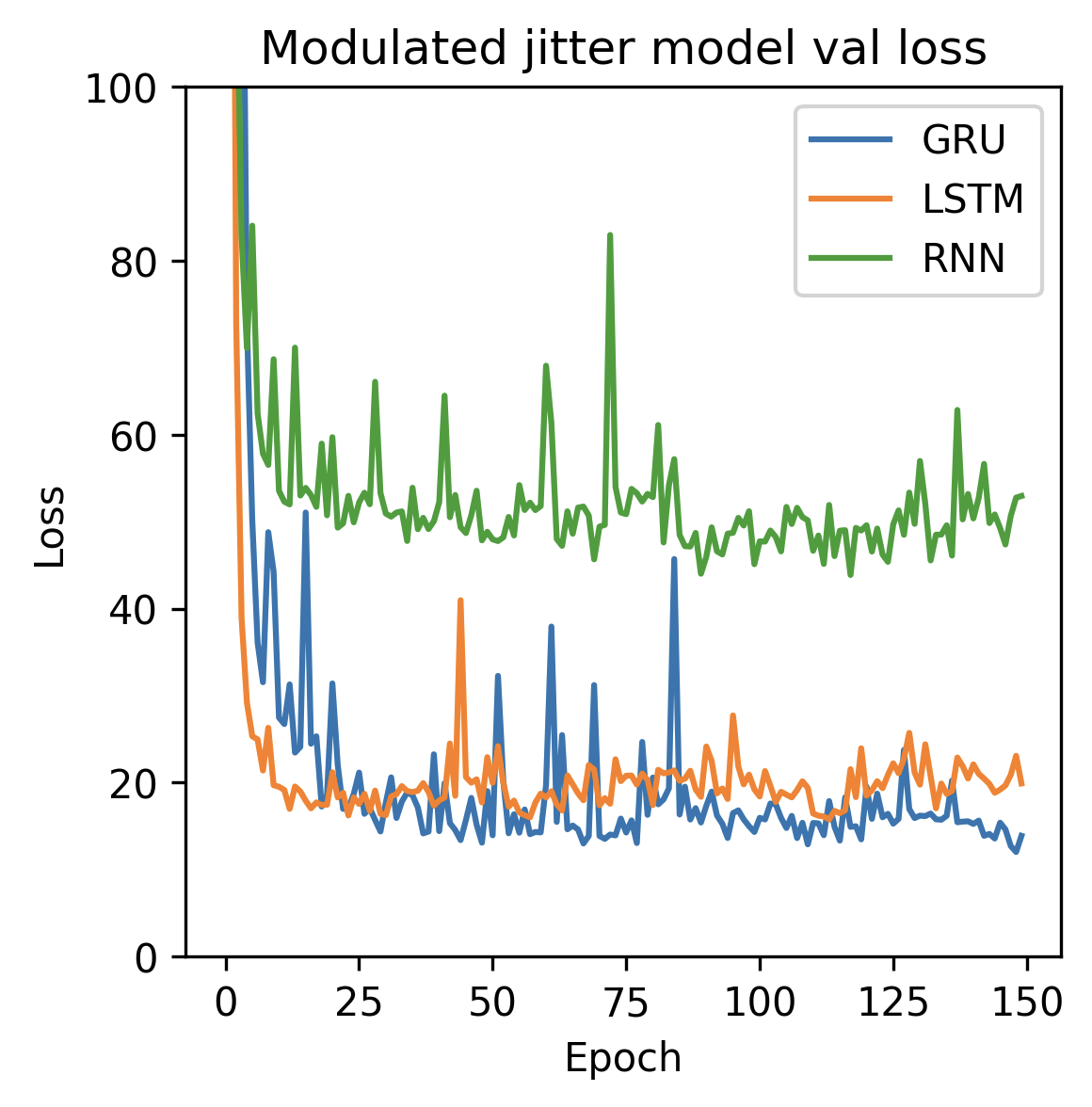}
   \caption{modul-jitter}
   \label{fig:moduljitter}
    \end{subfigure}
    \begin{subfigure}[b]{0.32\linewidth}
   \centering
   \includegraphics[width=\linewidth]{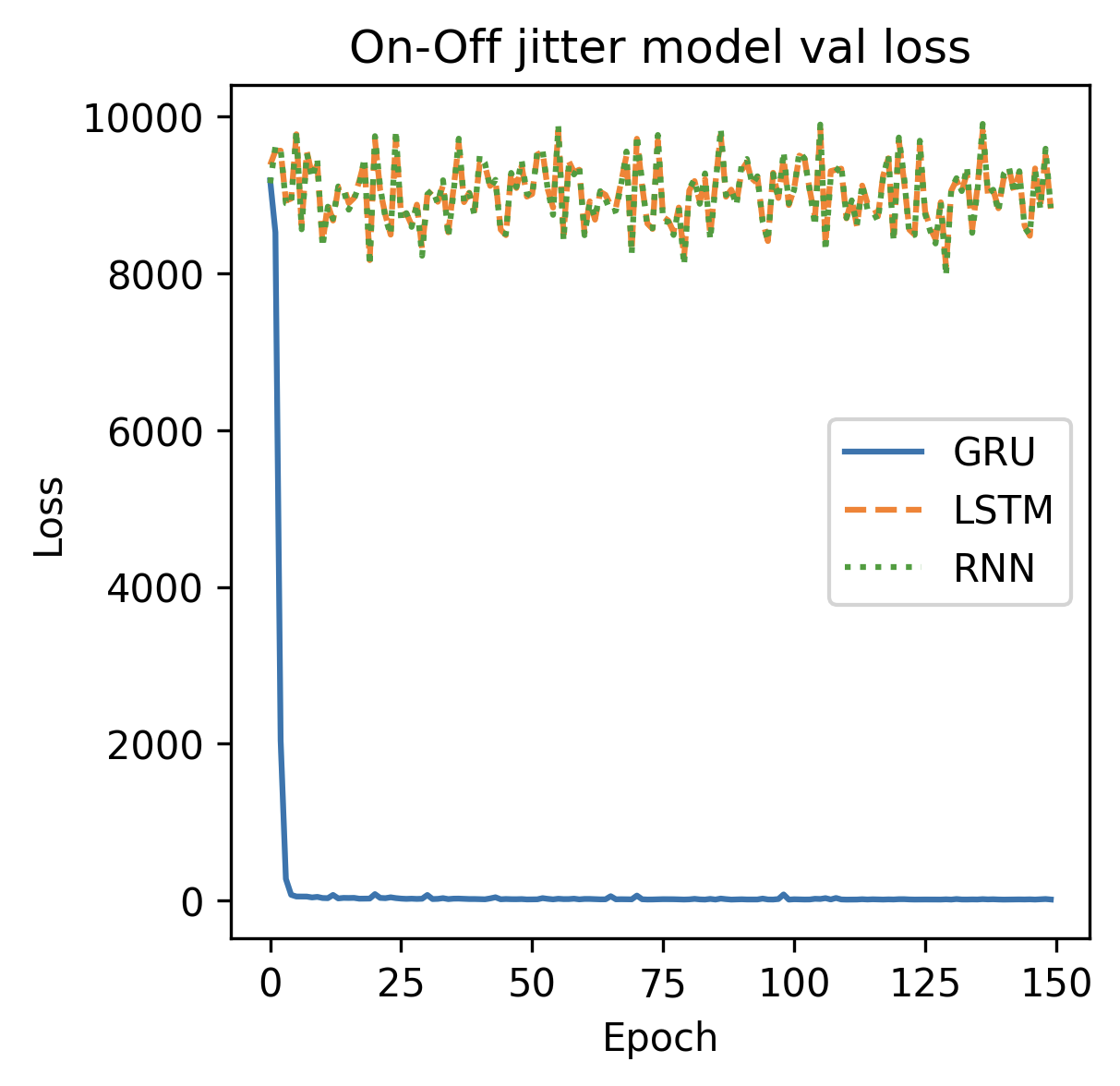}
   \caption{onoff-jitter}
   \label{fig:onoffjitter}
    \end{subfigure}
    \caption{Comparison of model validation loss per epoch for Traffic Models for jitter prediction}
    \label{fig:trafficjitter}
\end{figure}

For network topologies with different traffic models, LSTM architecture demonstrated superior performance in delay prediction across most scenarios, achieving the lowest MAPE values ranging from 2.16\% to 5.42\%. For Constant Bitrate traffic, LSTM achieved 2.21\% MAPE compared to RNN's 5.39\% and GRU's 3.76\%. However, the original paper's GRU implementation consistently performed better in jitter prediction, with MAE values as low as 0.008 for Constant Bitrate and Autocorrelated traffic. Both LSTM and GRU showed comparable performance in packet loss prediction, typically achieving MAE values around 0.004-0.005. The basic RNN model generally underperformed, particularly struggling with Modulated traffic (7.81\% MAPE) and All Multiplexed scenarios (12.29\% MAPE). A significant challenge emerged in the jitter prediction task, where validation loss curves shown in figure \ref{fig:trafficjitter} displayed inconsistent behavior across all models, indicating an area requiring further research and improvement.

\begin{table}[H]
\centering
\caption{Comparison of model metrics on test dataset for
Full Traffic Model topology for delay jitter and losses prediction}
\resizebox{0.49\textwidth}{!}{
\begin{tabular}{cc|ccc}
\cmidrule(r){1-5}
Traffic Model & Model & Delay MAPE & Jitter MAE & Losses MAE \\
\cmidrule(r){1-5}
\multirow{3}{*}{Constant Bitrate} 
& Paper GRU & 4.43\% & \textbf{0.008}  & 0.006 \\
& GRU Model & 3.76\% & 0.015  & 0.005 \\
& LSTM Model& \textbf{2.21\%} & 0.164  & \textbf{0.004} \\
& RNN Model & 5.39\% & 0.035  & 0.009 \\
\cmidrule(r){1-5}
\multirow{3}{*}{On/Off}
& Paper GRU & 2.90\% & \textbf{0.018}  & 0.006 \\
& GRU Model & \textbf{2.82\%} & 0.028  & 0.005 \\
& LSTM Model& 2.84\% & 0.272  & \textbf{0.004} \\
& RNN Model & 4.91\% & 0.269  & 0.008 \\
\cmidrule(r){1-5}
\multirow{3}{*}{Autocorrelated} 
& Paper GRU & 2.62\% & \textbf{0.008}  & 0.050 \\
& GRU Model & 2.68\% & 0.195  & \textbf{0.004} \\
& LSTM Model& \textbf{2.16\%} & 0.195  & 0.024 \\
& RNN Model & 4.38\% & 0.194  & 0.046 \\
\cmidrule(r){1-5}
\multirow{3}{*}{Modulated} 
& Paper GRU & \textbf{5.21\%} & \textbf{0.091}  & \textbf{0.010} \\
& GRU Model & 5.29\% & 0.099  & 0.011 \\
& LSTM Model& 5.42\% & 0.116  & 0.011 \\
& RNN Model & 7.81\% & 0.311  & 0.015 \\
\cmidrule(r){1-5}
\multirow{3}{*}{All Multiplexed} 
& Paper GRU & 4.71\% & \textbf{0.034}  & 0.005 \\
& GRU Model & 3.67\% & 0.052  & 0.005 \\
& LSTM Model& \textbf{3.49\%} & 0.503  & \textbf{0.004} \\
& RNN Model & 12.29\% & 0.185 & 0.009 \\
\cmidrule(r){1-5}
\end{tabular}
}
\label{tab:trafficmodel}
\end{table}

\begin{figure}[ht]
    \centering
    \begin{subfigure}[b]{0.32\linewidth}
   \centering
   \includegraphics[width=\linewidth]{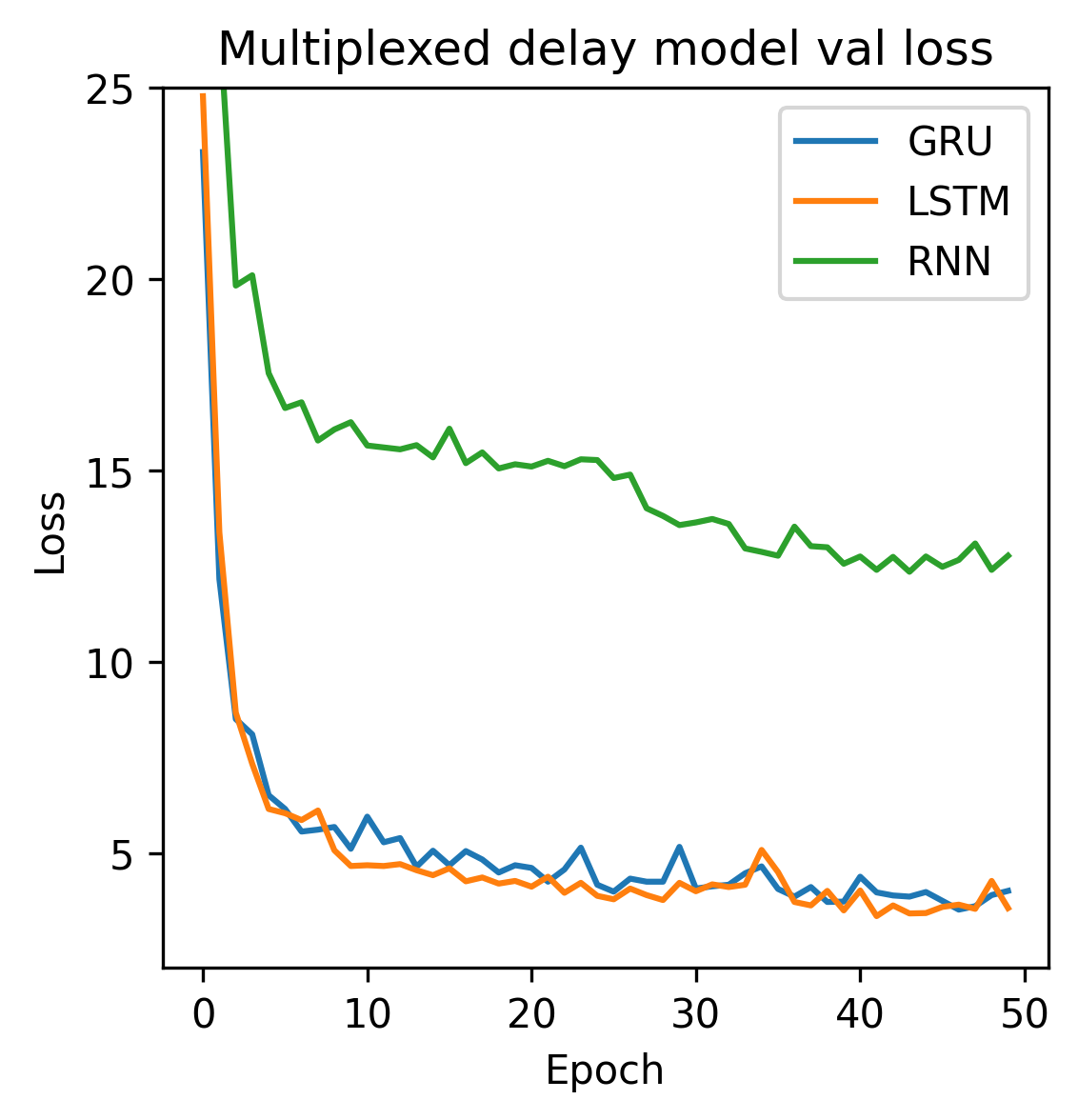}
   \caption{multi-delay}
   \label{fig:multidelay}
    \end{subfigure}
    \hfill
    \begin{subfigure}[b]{0.32\linewidth}
   \centering
   \includegraphics[width=\linewidth]{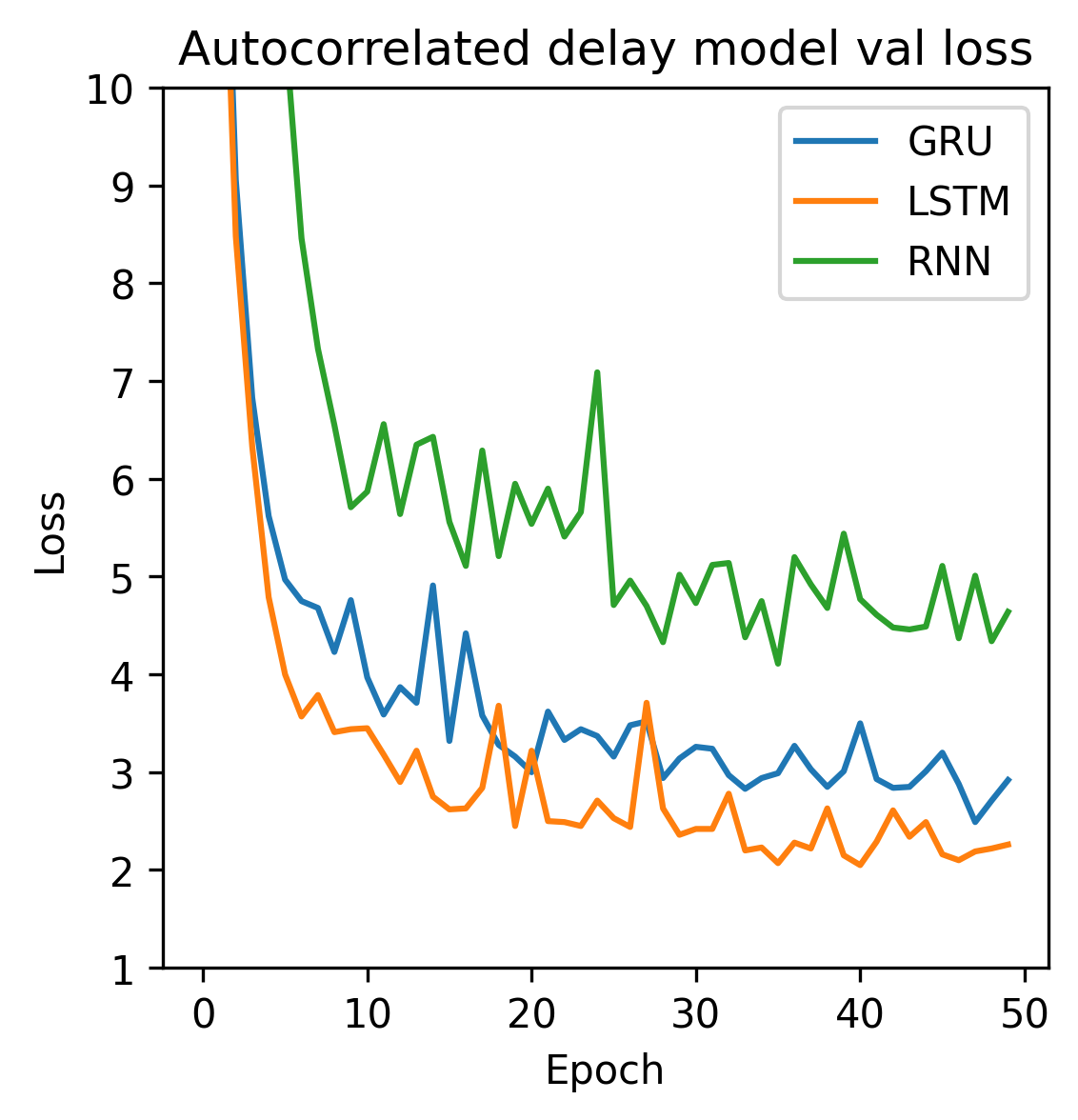}
   \caption{auto-delay}
   \label{fig:autodelay}
    \end{subfigure}
    \hfill
    \begin{subfigure}[b]{0.32\linewidth}
   \centering
   \includegraphics[width=\linewidth]{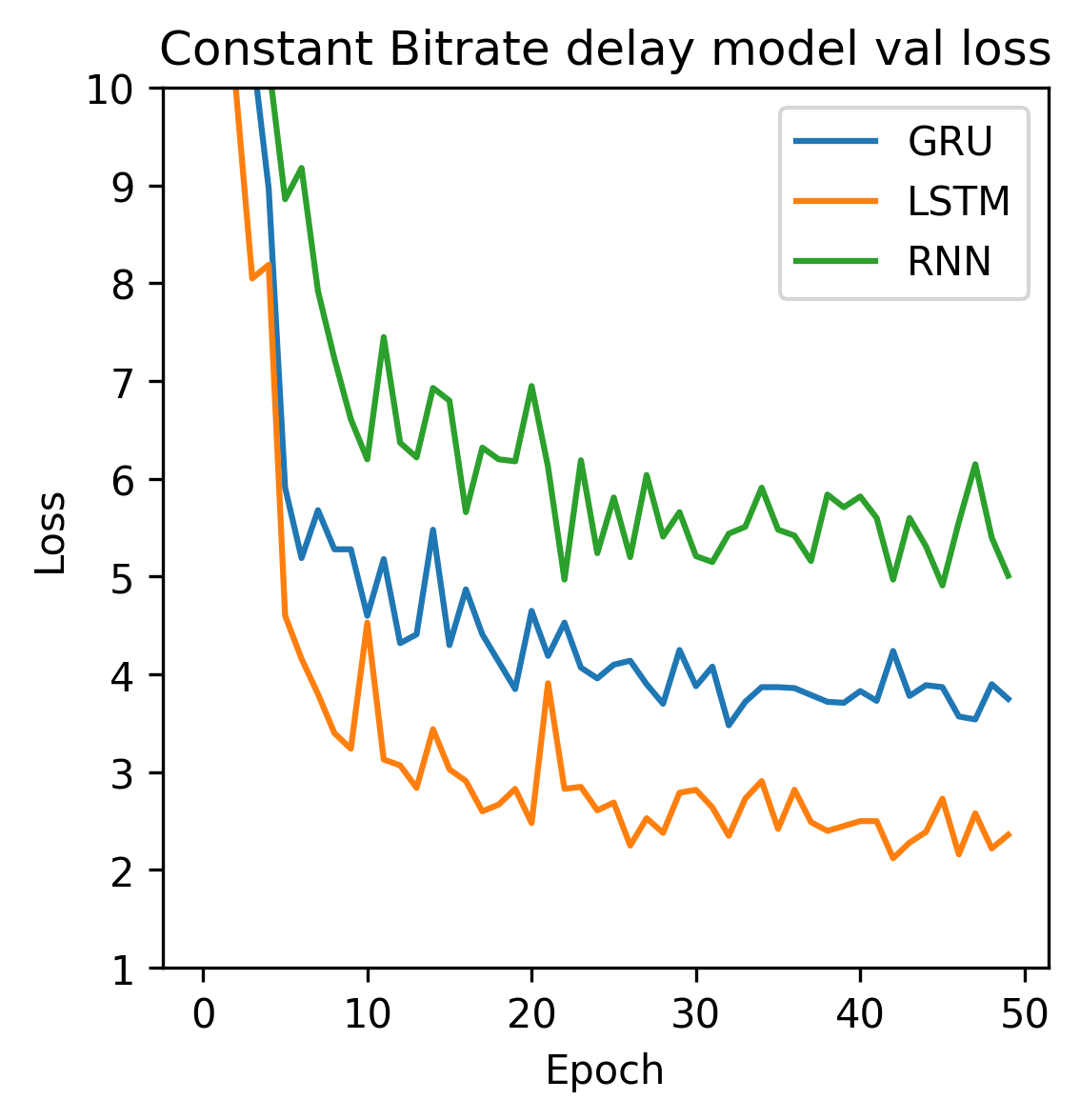}
   \caption{bit-delay}
   \label{fig:bitdelay}
    \end{subfigure}
    \hfill
    \begin{subfigure}[b]{0.32\linewidth}
   \centering
   \includegraphics[width=\linewidth]{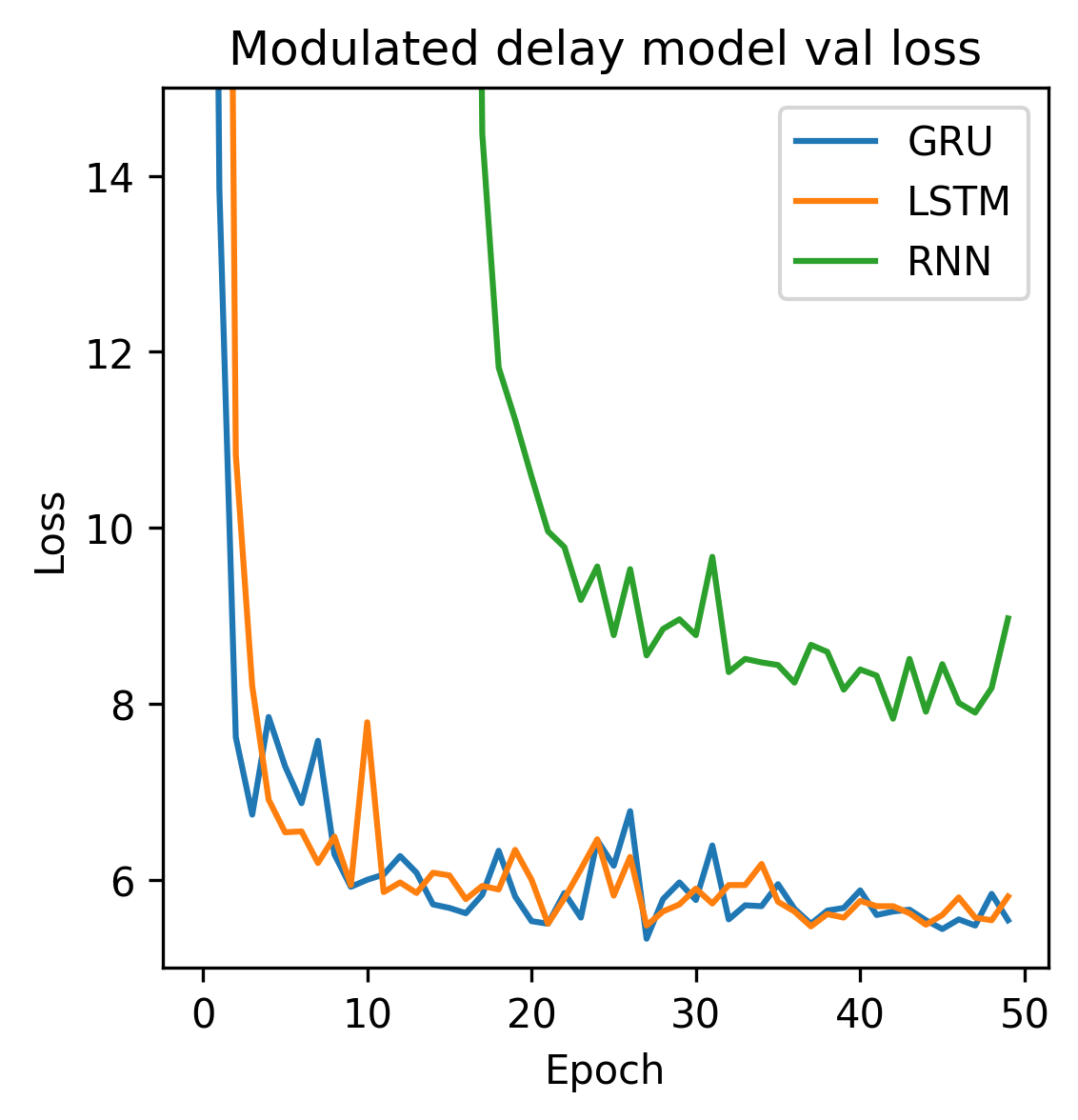}
   \caption{modul-delay}
   \label{fig:moduldelay}
    \end{subfigure}
    \begin{subfigure}[b]{0.32\linewidth}
   \centering
   \includegraphics[width=\linewidth]{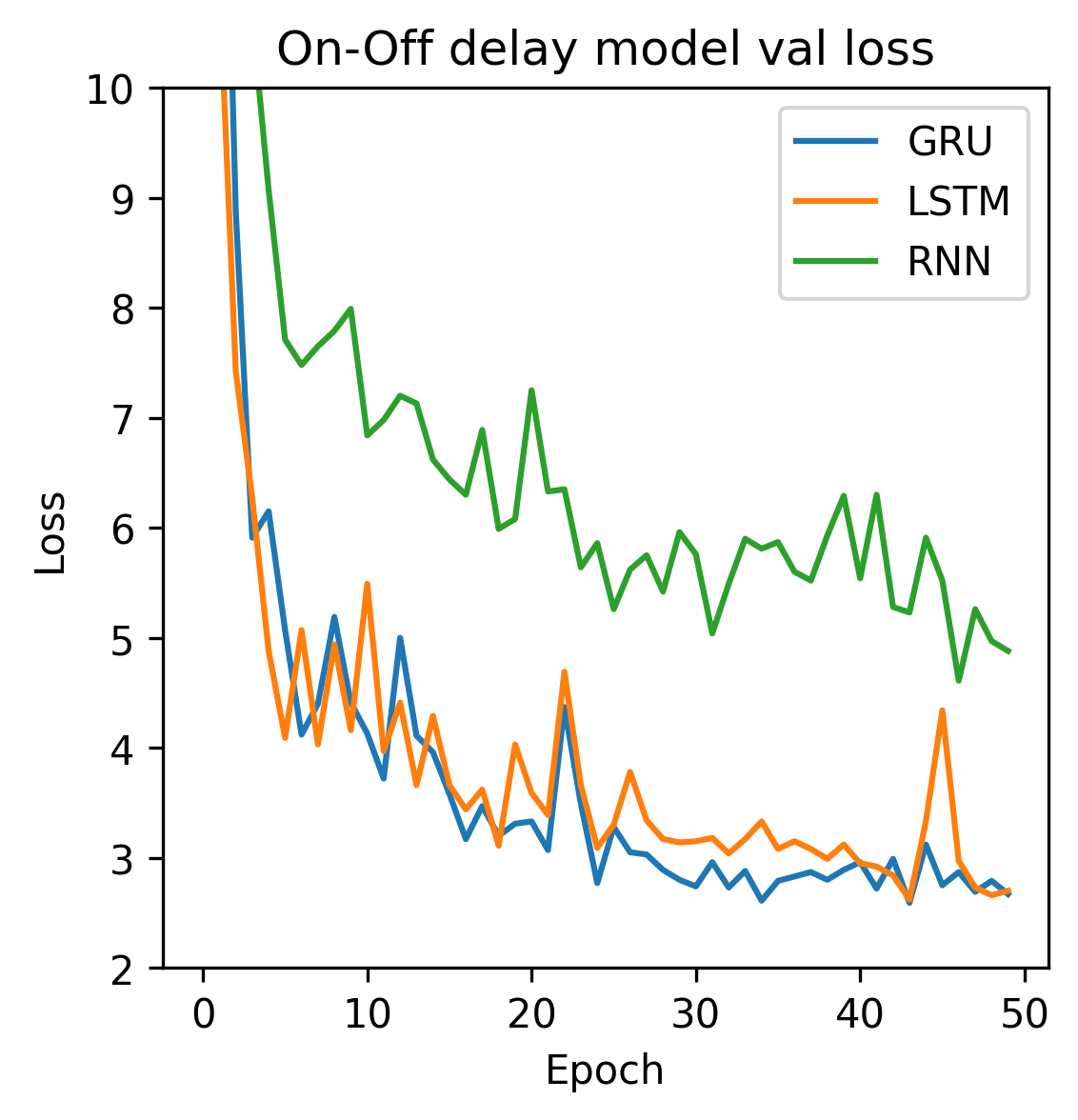}
   \caption{onoff-delay}
   \label{fig:onoffdelay}
    \end{subfigure}
    \caption{Comparison of model validation loss per epoch for Traffic Models for delay prediction}
    \label{fig:trafficdelay}
\end{figure}

\begin{figure}[ht]
    \centering
    \begin{subfigure}[b]{0.32\linewidth}
   \centering
   \includegraphics[width=\linewidth]{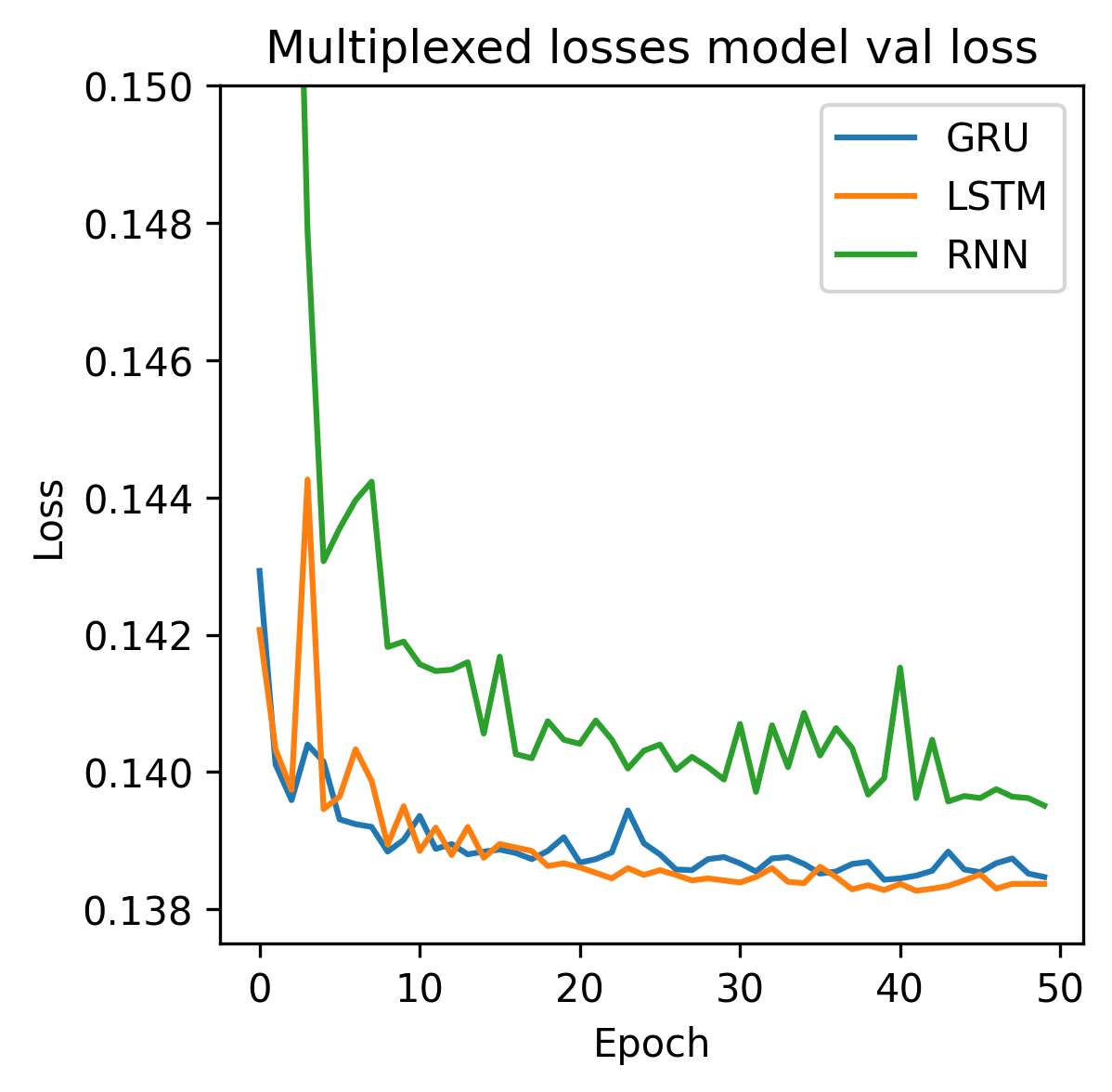}
   \caption{multi-loss}
   \label{fig:multiloss}
    \end{subfigure}
    \hfill
    \begin{subfigure}[b]{0.32\linewidth}
   \centering
   \includegraphics[width=\linewidth]{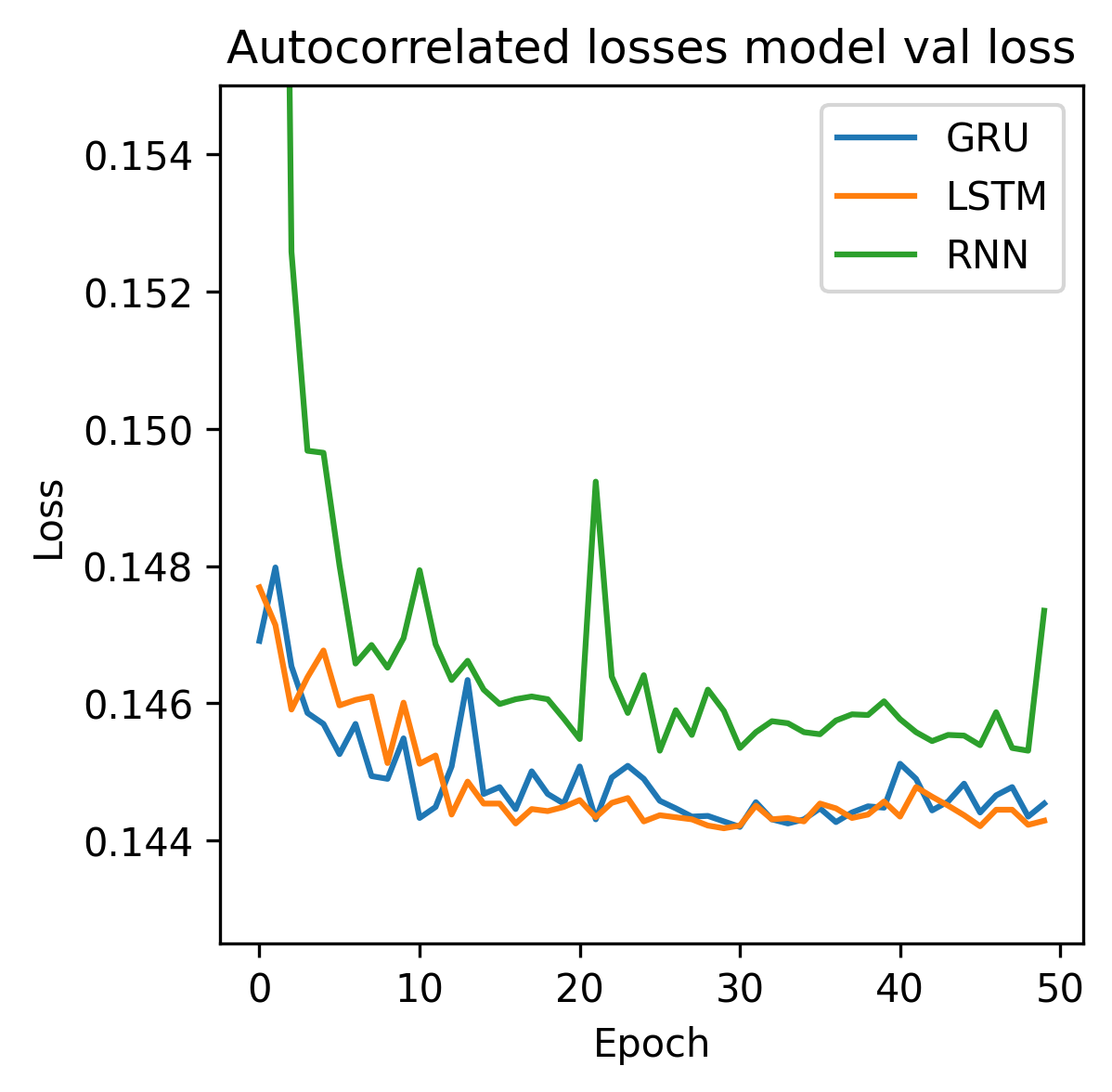}
   \caption{auto-loss}
   \label{fig:autoloss}
    \end{subfigure}
    \hfill
    \begin{subfigure}[b]{0.32\linewidth}
   \centering
   \includegraphics[width=\linewidth]{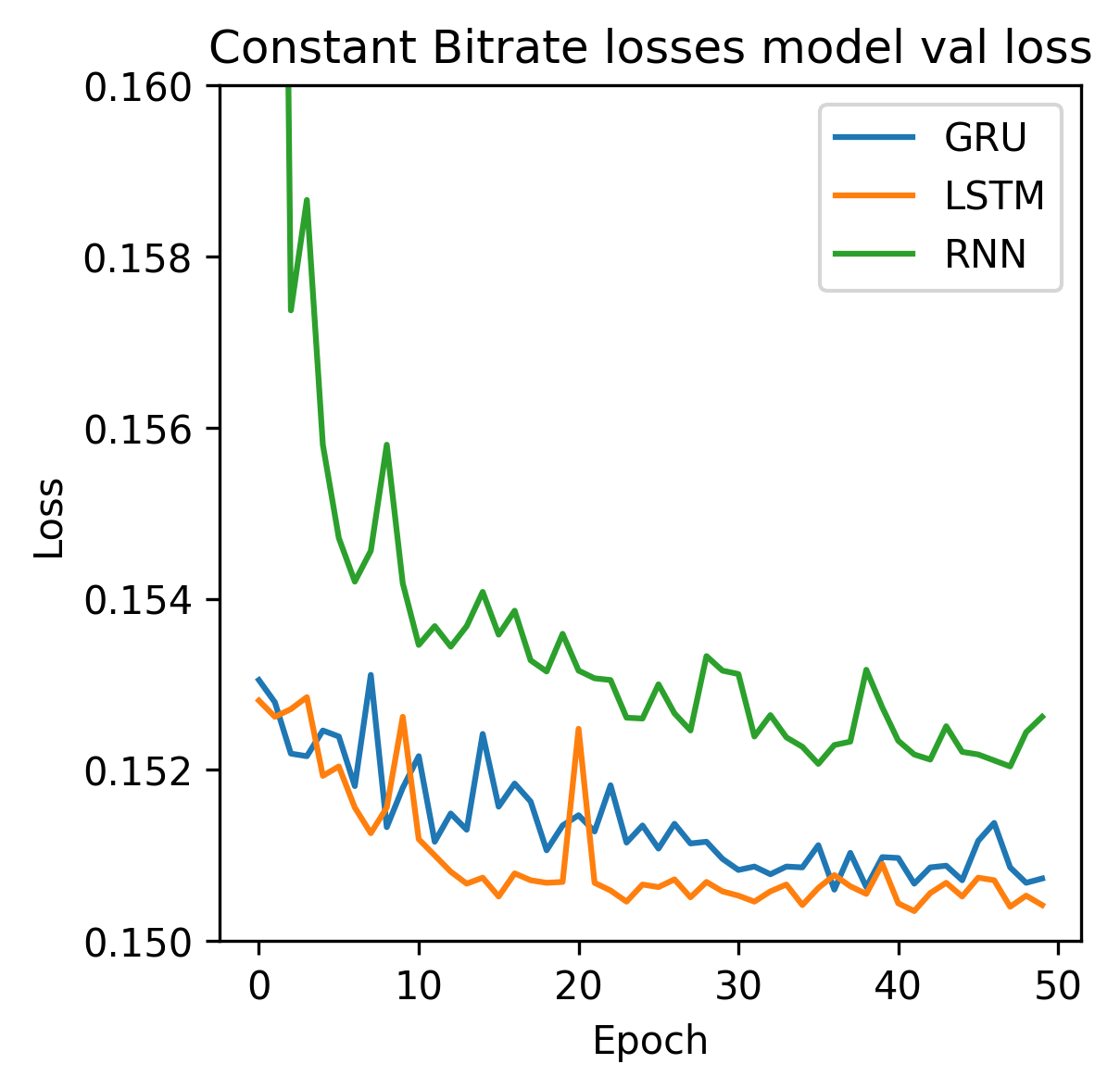}
   \caption{bit-loss}
   \label{fig:bitloss}
    \end{subfigure}
    \hfill
    \begin{subfigure}[b]{0.32\linewidth}
   \centering
   \includegraphics[width=\linewidth]{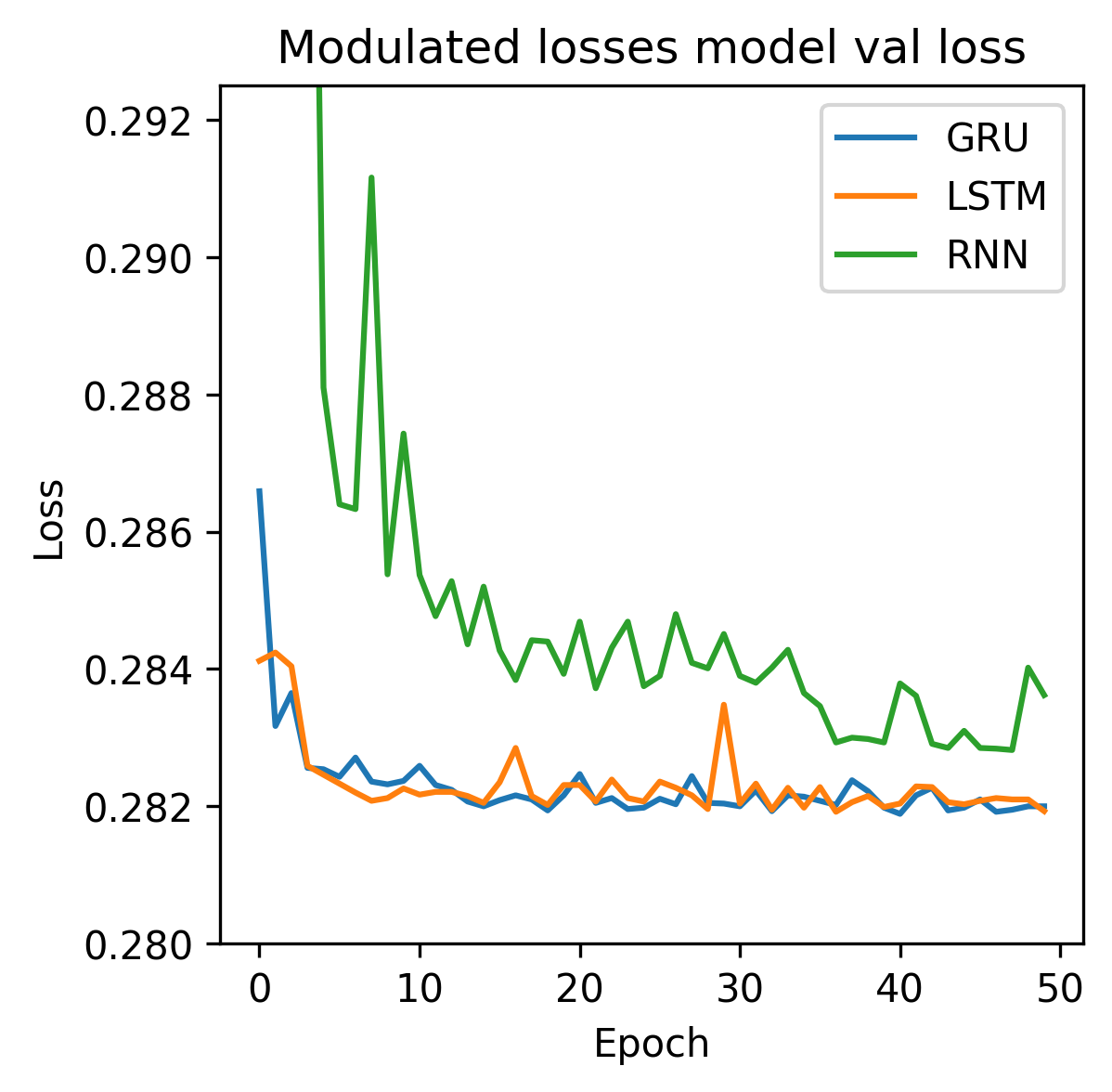}
   \caption{modul-loss}
   \label{fig:modulloss}
    \end{subfigure}
    \begin{subfigure}[b]{0.32\linewidth}
   \centering
   \includegraphics[width=\linewidth]{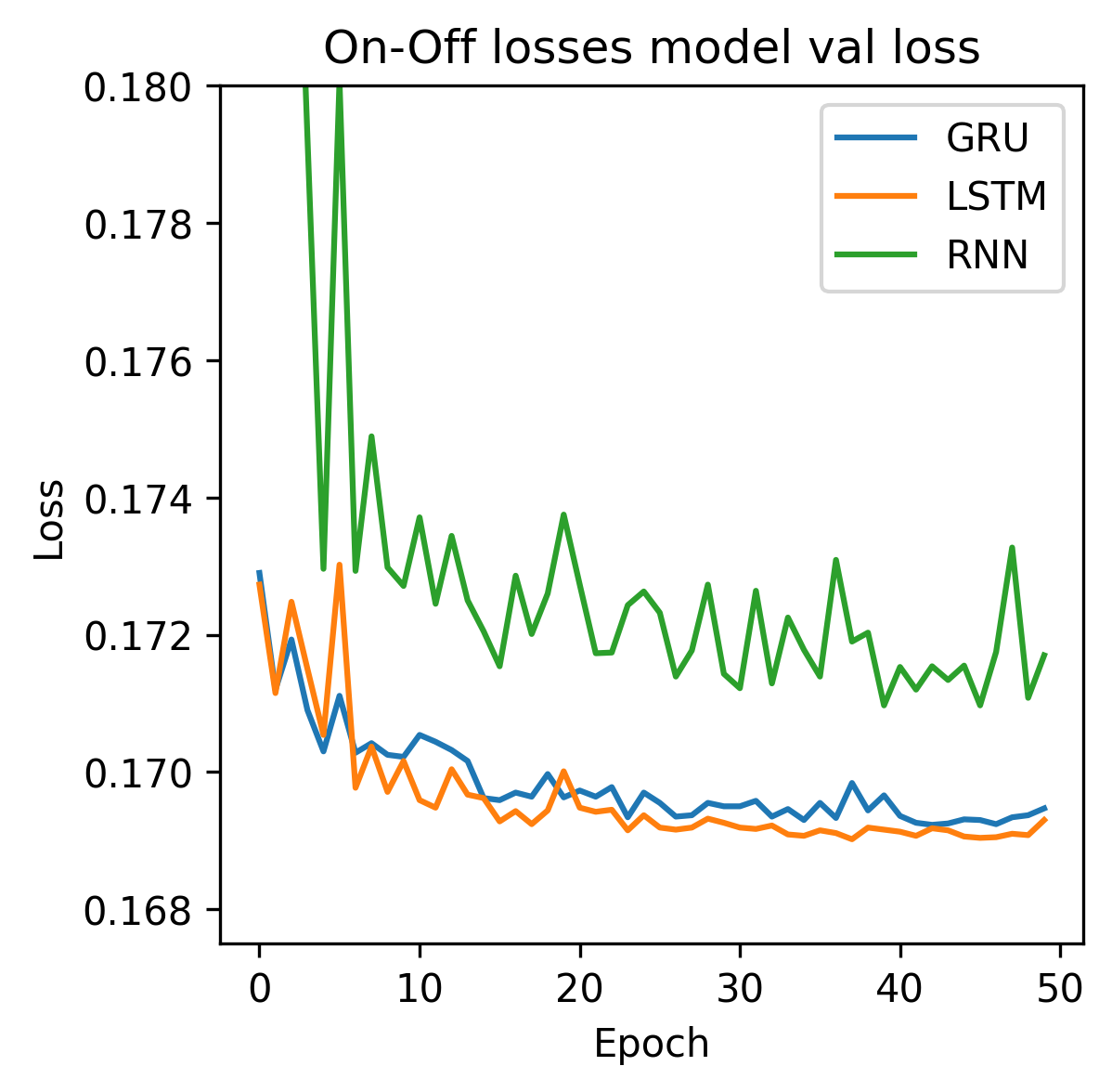}
   \caption{onoff-loss}
   \label{fig:onoffloss}
    \end{subfigure}
    \caption{Comparison of model validation loss per epoch for Traffic Models for packet loss prediction}
    \label{fig:trafficloss}
\end{figure}

\section{Discussion}

This work presents a comprehensive analysis of different recurrent neural architectures within the RouteNet-Fermi framework for network performance prediction. Our implementation and evaluation of RNN, LSTM, and GRU variants revealed several key findings. The LSTM architecture demonstrated superior performance across most scenarios, particularly in delay prediction. GRU cells maintained a balanced trade-off between computational efficiency and prediction accuracy, matching or exceeding the original paper's results in several cases. While the basic RNN showed competitive performance in simpler scenarios, it struggled with complex traffic patterns and larger networks. Model performance varied significantly across different network tasks, with LSTM showing particular strength in capturing temporal dependencies critical for jitter prediction. Despite the promising results, our implementation faces several key limitations which include computational resources, dataset constraints, and model complexity trade-offs. Training on CPU limited our ability to experiment with larger hidden state dimensions and batch sizes, potentially impacting model performance. While we covered various network scenarios, our training data may not fully represent all possible real-world traffic patterns and network configurations. Also for traffic model jitter prediction task, the training struggled compared to the paper. The LSTM variant, while more accurate, introduces higher computational overhead compared to simpler RNN architectures, affecting inference time particularly in larger networks. Performance degradation was observed when scaling to networks significantly larger than training topologies, especially with the basic RNN implementation. These findings advance our understanding of recurrent architectures in network modeling and provide valuable insights for future GNN-based network performance prediction systems. The framework's modular design enables further experimentation with different cell types, contributing to the ongoing evolution of data-driven network modeling approaches.

\appendix

\section{Appendix}

Here all the results from the tables are plotted.

\begin{figure}[H]
    \centering
    \includegraphics[width=0.9\linewidth]{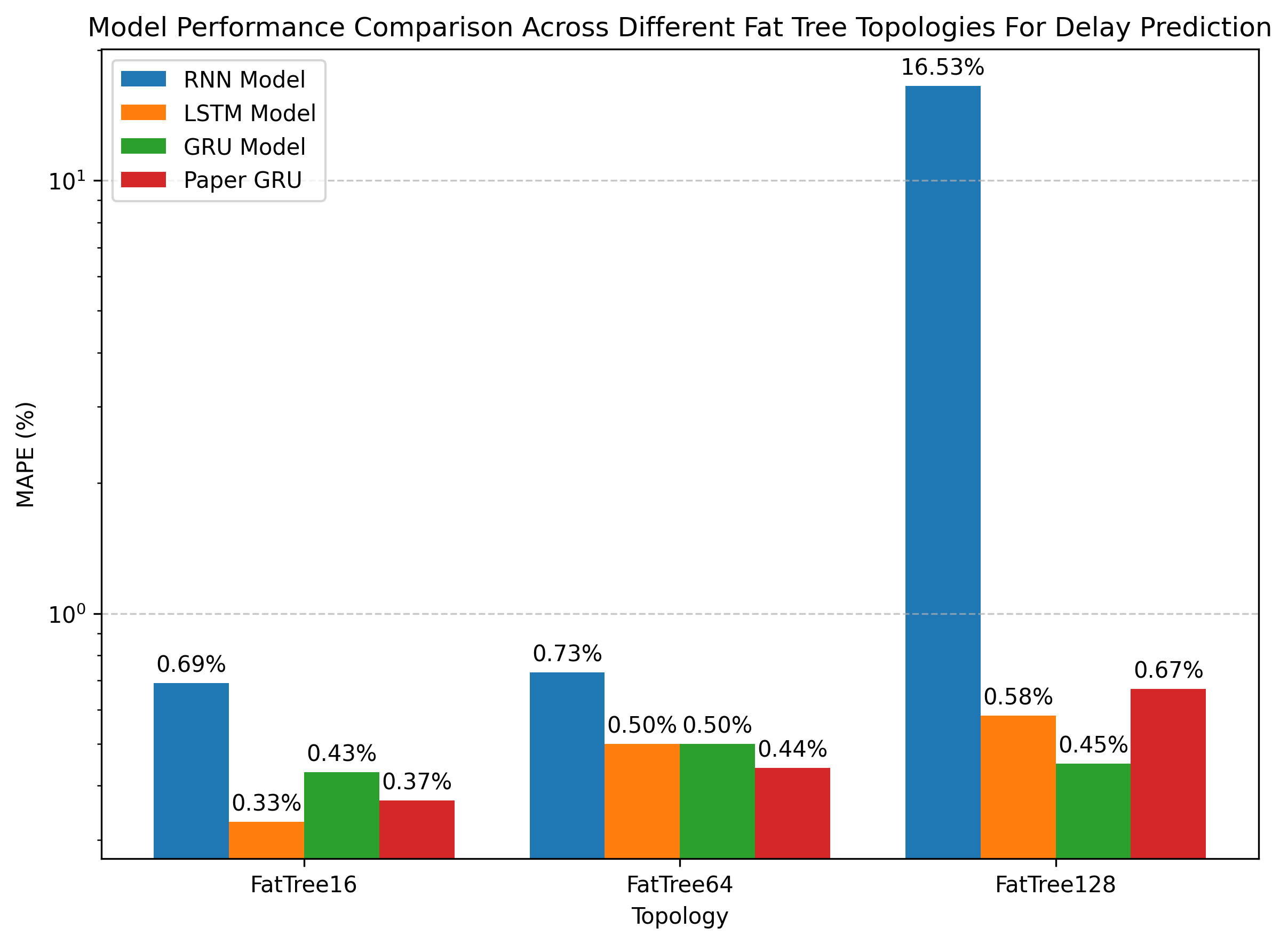}
    \label{fig:fattreebar}
\end{figure}

\begin{figure}[H]
    \centering
    \includegraphics[width=0.9\linewidth]{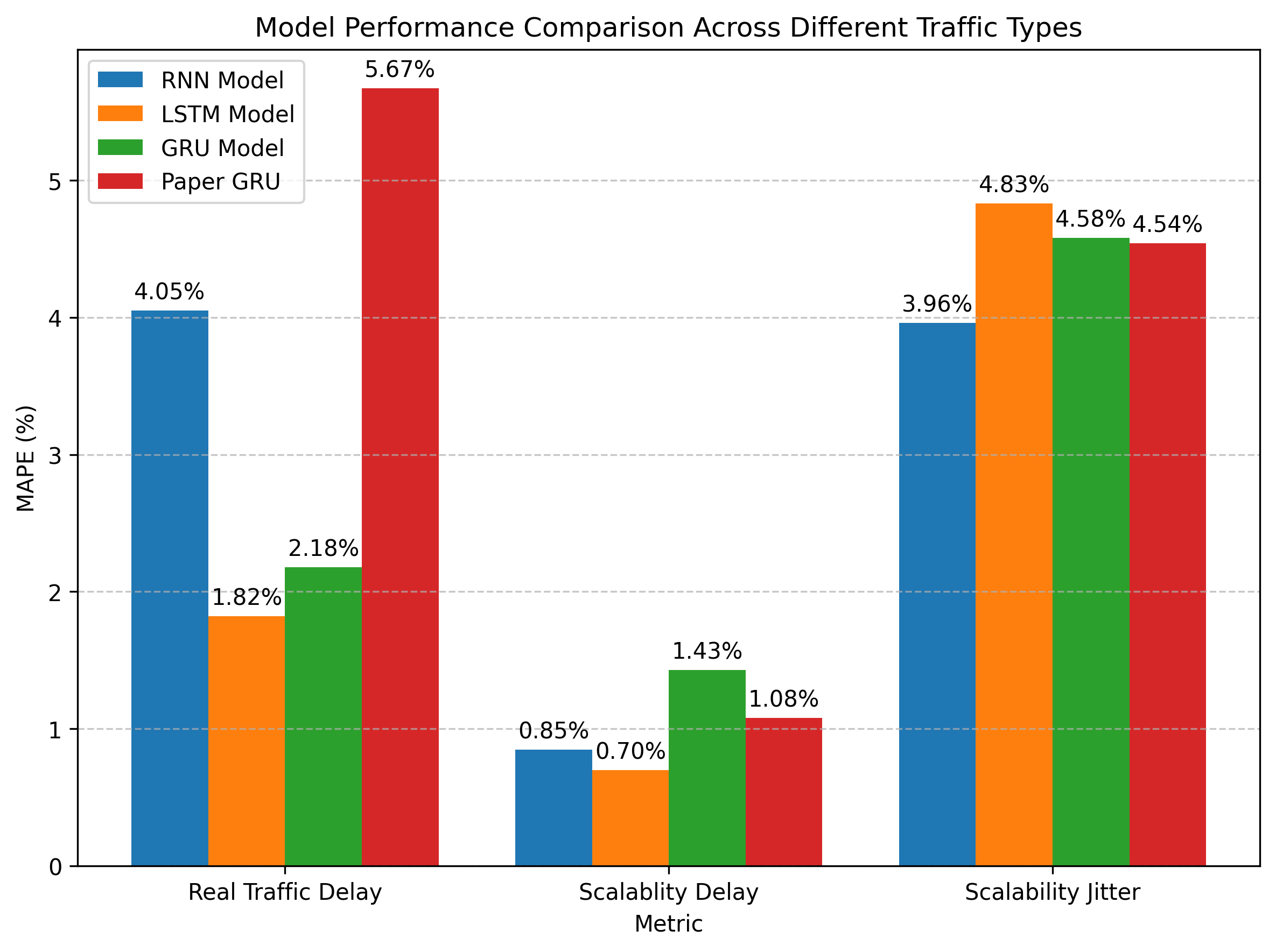}
    \label{fig:scalebar}
\end{figure}

\begin{figure}[H]
    \centering
    \includegraphics[width=0.9\linewidth]{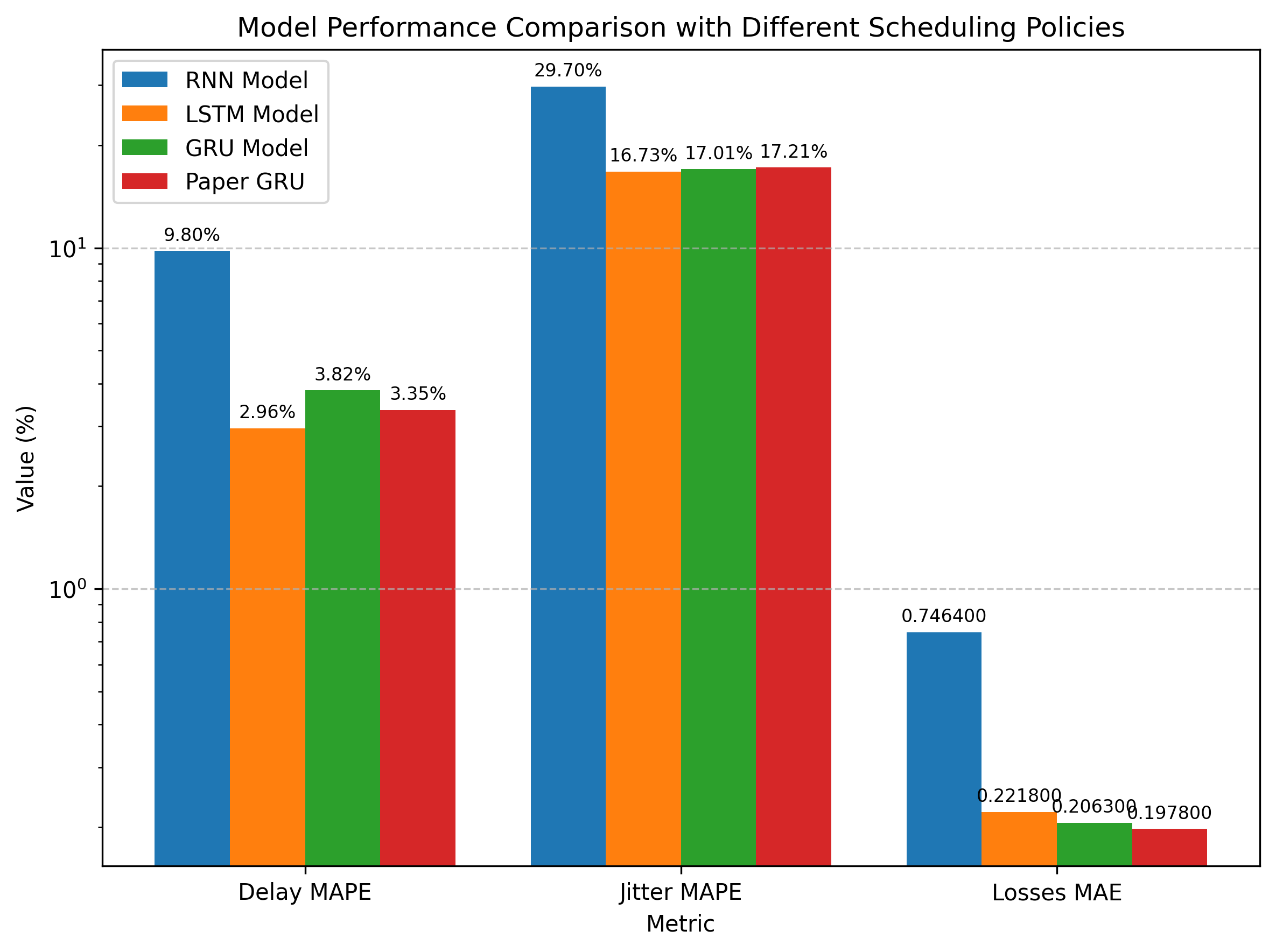}
    \label{fig:schedulebar}
\end{figure}

\begin{figure}[H]
    \centering
    \includegraphics[width=0.9\linewidth]{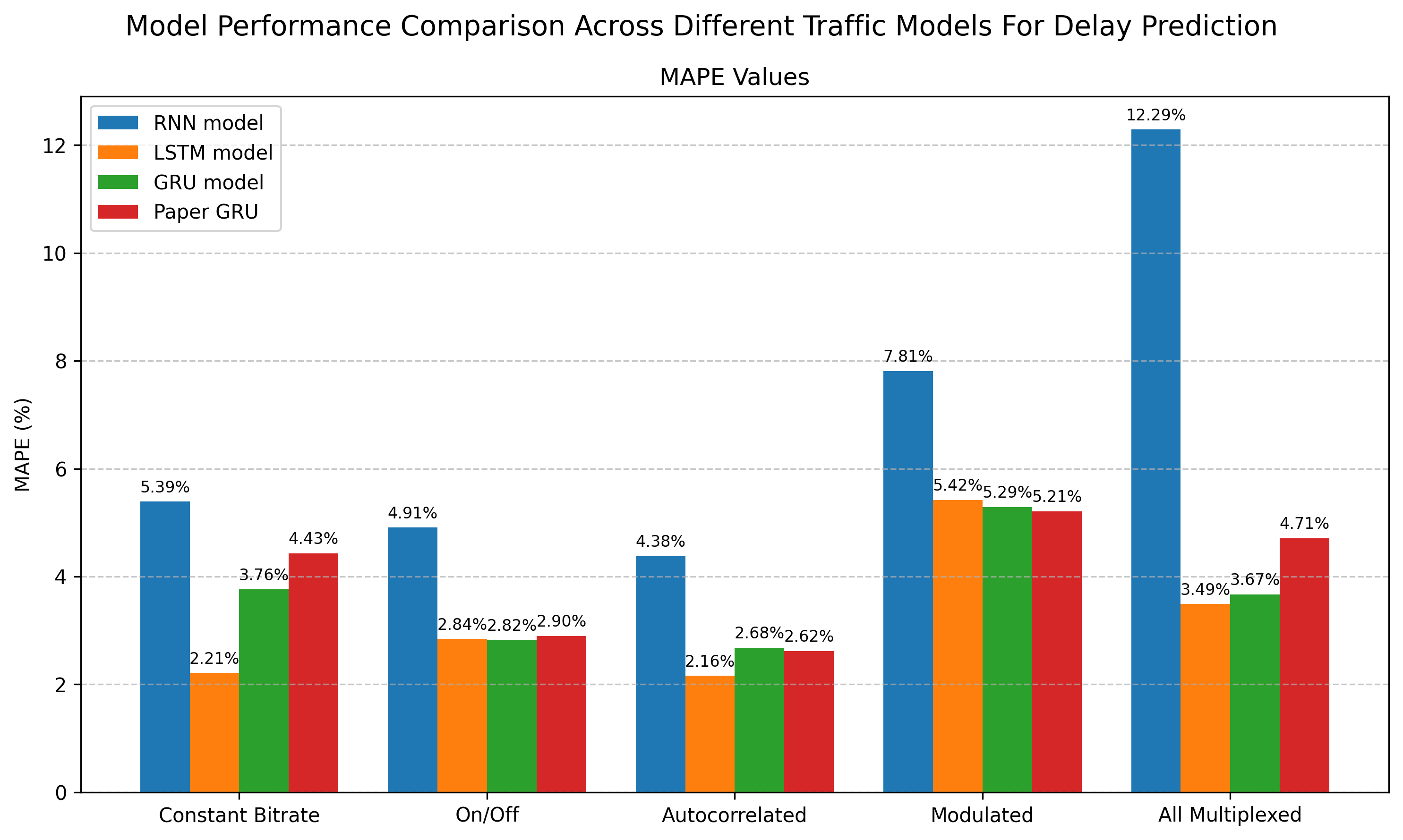}
    \label{fig:trafficdelaybar}
\end{figure}

\begin{figure}[H]
    \centering
    \includegraphics[width=0.9\linewidth]{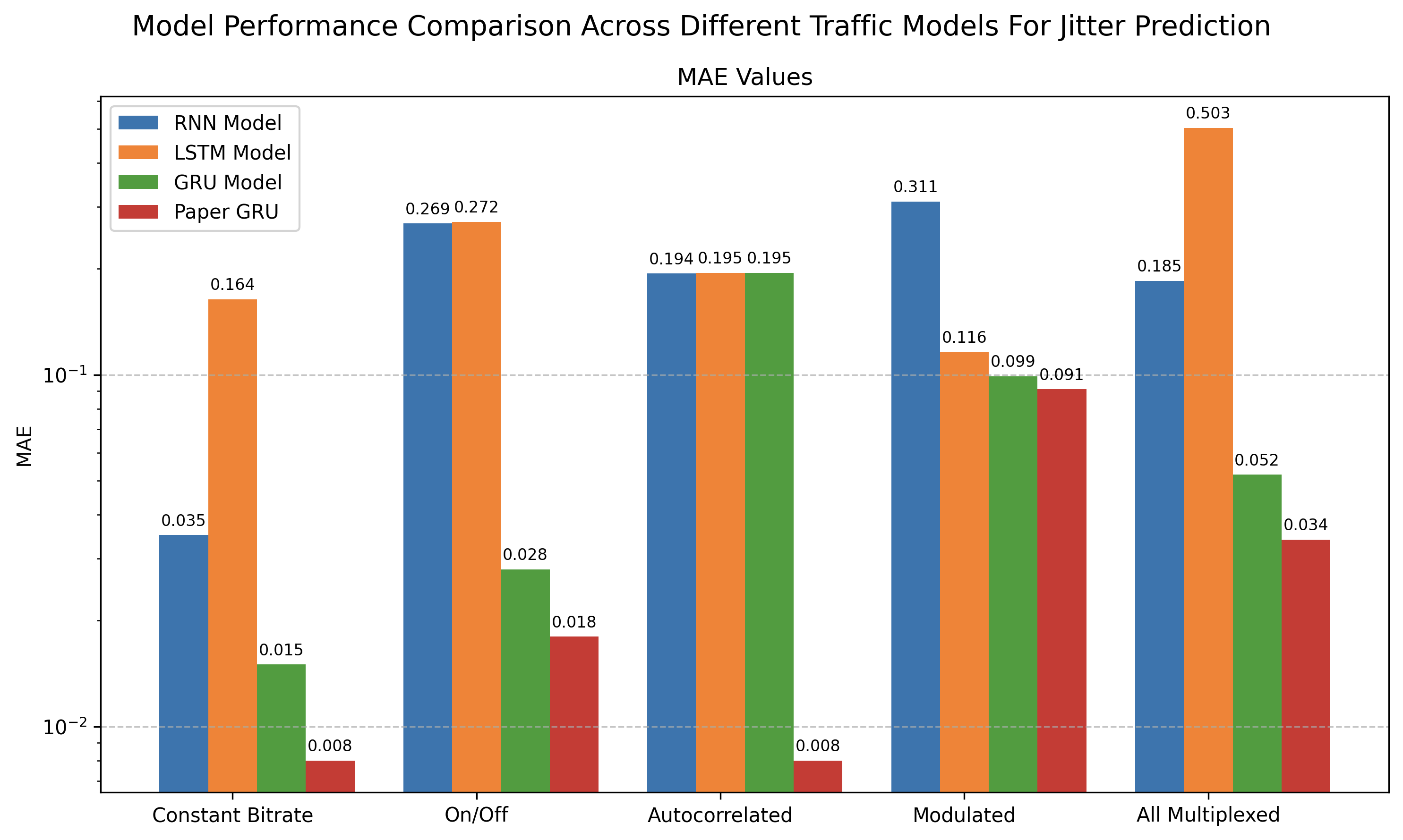}
    \label{fig:trafficjitterbar}
\end{figure}

\begin{figure}[H]
    \centering
    \includegraphics[width=0.9\linewidth]{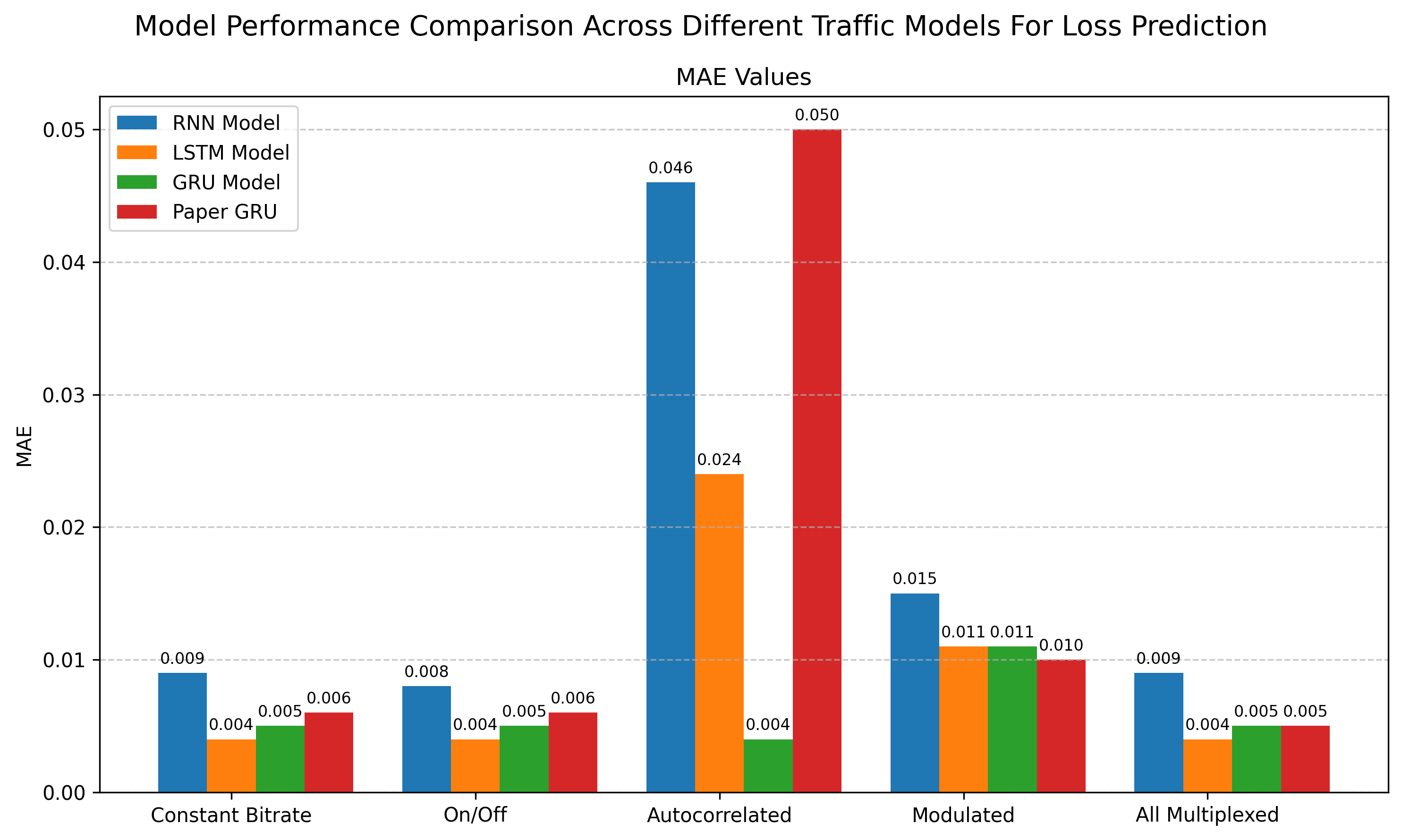}
    \label{fig:trafficlossbar}
\end{figure}

\end{document}